\documentclass[
    reprint,
    amsmath,
    amssymb,
    aps,prd,
    floatfix,
    nofootinbib,
    superscriptaddress
    ]{revtex4-2}

\usepackage{graphicx} 
\usepackage{xcolor} 
\usepackage[
    colorlinks=true,
    linkcolor=blue,
    citecolor=blue,
    urlcolor=blue
]{hyperref} 
\usepackage{orcidlink}
\usepackage{enumitem} 

\setlist[enumerate,1]{
    label={\!\!\alph*)},
    leftmargin=*,
    nosep
}

\setlist[itemize,1]{
    leftmargin=*,
    nosep
}

\usepackage[left=1.9cm, 
            right=1.4cm, 
            top=1.8cm, 
            bottom=1.5cm]{geometry}
\usepackage{multirow}  

\graphicspath{{./Figures/}}

\renewcommand{\v}[1]{\boldsymbol{#1}}
\renewcommand{\prl}{\parallel}

\newcommand{\pbr}[1]{\left(#1\right)}

\newcommand{\lrangle}[1]{\langle #1 \rangle}

\newcommand{\entry}[1]{\vspace{1.0ex}\noindent\textit{#1}\vspace{.5ex}}

\renewcommand{\rm}[1]{\mathrm{#1}}
\newcommand{\dm}[1]{\mathrm{~#1}}
\newcommand{\rad}{\rm{rad}}
\newcommand{\pc}{\rm{pc}}
\newcommand{\uG}{\rm{\mu G}}
\newcommand{\PeV}{\rm{PeV}}
\newcommand{\TeV}{\rm{TeV}}
\newcommand{\GeV}{\rm{GeV}}
\newcommand{\keV}{\rm{keV}}

\newcommand{\Sa}{S_\alpha}
\newcommand{\Ca}{C_\alpha}
\newcommand{\Ta}{T_\alpha}

\newcommand{\p}{\partial}

\begin{document}

    \title{
    LHAASO's Peanut source is a laboratory
    for studying magnetic fields in the Milky Way
}

\author{Maksim~Kleimenov\,\orcidlink{0009-0001-5593-3401}}
\affiliation{Universit\'e Paris Cit\'e, CNRS, AstroParticule et Cosmologie, F-75013 Paris, France}
\affiliation{Institut Polytechnique de Paris, Palaiseau, \^Ile-de-France, France}

\author{Yuan Li\,\orcidlink{0009-0003-4873-6770}}
\affiliation{Tsung-Dao Lee Institute, Shanghai Jiao Tong University, Shanghai, P.R. China}

\author{Gwenael Giacinti}
\affiliation{Universit\'e Paris Cit\'e, CNRS, AstroParticule et Cosmologie, F-75013 Paris, France}

\author{Dmitri Semikoz}
\email{
    mkleimenov@apc.in2p3.fr,
    yuanlss17@sjtu.edu.cn\\
    ggiacinti@apc.in2p3.fr,
    dmitri.semikoz@apc.univ-paris7.fr}
\affiliation{Universit\'e Paris Cit\'e, CNRS, AstroParticule et Cosmologie, F-75013 Paris, France}

\date{November 11, 2026}

\begin{abstract}
    Recently, the LHAASO Collaboration reported gamma-ray emission from the Peanut-shaped source (Peanut), which extends over $4.5^\circ$ parallel to the Galactic plane at $b=-17.5^\circ$. The source is associated with the millisecond pulsar PSR~J0218+4232, located at the eastern edge of the Peanut, at a distance of $3\dm{kpc}$ from the Earth and $0.9\dm{kpc}$ from the Galactic plane. Its location and bright gamma-ray emission above $25~\TeV$ make it a natural laboratory for studying the Galactic magnetic field (GMF) in the transition region between the Galactic plane and halo. We interpret the observed emission as inverse-Compton scattering of CMB photons by electrons propagating along the large-scale GMF, accounting for turbulence through two models: Kolmogorov and Goldreich-Sridhar~(GS). These models correspond to diffusive and ballistic particle propagation, respectively. While the GS case naturally provides conditions for one-sided gamma-ray emission, in the Kolmogorov case assumptions of one-sided particle injection and magnetic mirror at the source are necessary. We find that, in the Kolmogorov scenario, the Peanut's hotspots are likely associated with magnetic bottling, whereas in the GS scenario, they are related to magnetic field line geometry. Both models yield constraints on the GMF at the location of the Peanut.
\end{abstract}
    \maketitle

    \section{Introduction}

Extended gamma-ray sources may serve as probes for interstellar turbulent magnetic fields (MF), as demonstrated by theoretical studies~\cite{Giacinti2012, 2013GG}. Most of high-energy sources lie in the Galactic plane and show a morphology close to spherical~\cite{Cao2024}. Particle propagation for these objects, starting from Geminga and Monogem halos~\cite{HAWC2017}, is modeled with isotropic diffusion. With the accumulation of gamma-ray data, the LHAASO observatory reports the detection of an increasing number of asymmetric sources at the highest-energies ($> 25~\TeV$, e.g.~\cite{Cao2026b}). Among these sources, pulsar wind nebulae (PWNe) are of particular interest for characterization of the Galactic magnetic field (GMF).

The first LHAASO catalog revealed 12 point sources outside of the Galactic plane~\cite{Cao2024}. Three of them were located in one region at $b=-17.5^\circ$, aligned with the Galactic plane, and had a similar flux described with a power-law with a cutoff. All three sources appeared among a new population of PeVatrons with very significant flux above $100~\TeV$. 

Recently, the LHAASO collaboration published a paper proving that these three sources belong to a single elongated $4.5^\circ$ gamma-ray structure at $b=-17.5^\circ$ parallel to the Galactic plane~\cite{Cao2026}. This structure comprises three emission regions: a strip and two point-like sources, which earned the source a nickname ``Peanut''. So far, there is only one known candidate object able power the observed gamma-ray emission, the luminous millisecond pulsar PSR J0218+4238, located at 3 kpc from Earth and at about 0.9 kpc from the Galactic plane~\cite{Verbiest2014}. At this distance, the gamma-ray emission has a projected length above 200 pc, which makes this source an ideal laboratory for studying the magnetic field of the Galactic halo.

The GMF remains poorly understood despite decades of study. Existing models include several large-scale components: a disk field associated with the Galactic disk and spiral arms, a poloidal field in the central Galaxy, and a toroidal field extending above and below the Galactic  plane~~\cite{1997A&A...322...98H,Han:1999vi,Han:2006ci,Pshirkov:2011um,Jansson:2012pc,Unger:2023lob,Han2024,Korochkin:2024yit}. The detailed structure of these components remains uncertain in many regions, particularly in the Galactic halo.

The turbulent MF in the Galactic halo remains even less well characterized. State-of-the-art models~\cite{Jansson:2012rt,Unger:2026yau} provide estimates of its amplitude and coherence length, but as these estimates are generally integrated along the line of sight (LOS), they offer limited information about the local field structure. Extended gamma-ray emission around Galactic sources may provide a complementary probe of the GMF, potentially revealing its local structure through the propagation of high-energy particles~\cite{2013GG}.

In the present work, we model gamma-ray emission from LHAASO's Peanut source assuming inverse Compton (IC) scattering of CMB photons by high-energy electrons injected by PSR~J0218+4238. Injected electrons are transported in the combination of the regular and turbulent MF. For the latter, we consider two cases: Kolmogorov~\cite{Kolmogorov1941} and Goldreich-Sridhar (GS) turbulence~\cite{GoldreichSridhar1995}, which represent two limiting scenarios of diffusion and ballistic propagation, respectively.

This paper is structured as follows. In~\autoref{section:psr}, we introduce the Peanut source, PSR~J0218+4232, and assess the physical and energy scales of the Peanut region. In~\autoref{sec:GMF}, we discuss the Galactic magnetic-field structure and the two turbulence models considered in this work. In~\autoref{sec:diffusion} and~\autoref{sec:ballistic}, we investigate the diffusive and ballistic scenarios, respectively, assuming Kolmogorov and GS turbulence. For each scenario, we first derive analytical constraints and then perform simulations to assess its consistency with observations. In~\autoref{sec:discussion}, we discuss the results, their implications for Galactic magnetic-field models, and the challenges associated with interpreting each scenario. We summarize our findings in~\autoref{sec:conclusion}.

\section{The Peanut and PSR~J0218+4232}
\label{section:psr}

\begin{figure}[t!]
    \centering
    \includegraphics[width=\linewidth]{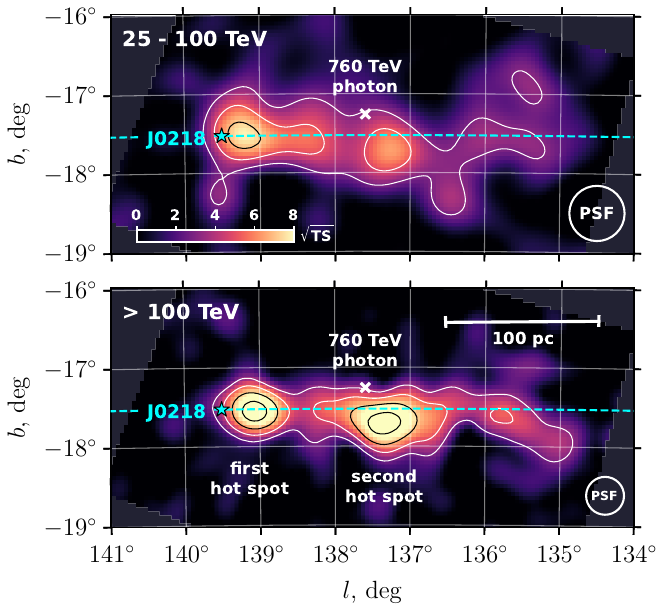}
    \caption{
    The Peanut source significance maps in the Galactic coordinates as seen by LHAASO~\cite{Cao2026} in two energy bins: $25 < \epsilon < 100~\TeV$ and $>100~\TeV$. Physical scales are derived assuming $d=3\dm{kpc}$. The figures  share the same color map; contours correspond to $\sqrt{TS}$ of 3, 5, 7, and 9. A cyan star shows the position of the pulsar J0218~\cite{Navarro1995}. The position of the highest-energy $760_{-40}^{+60}~\TeV$ Peanut's photon is shown with a white cross.  Rotation of the LHAASO data is performed with~\texttt{astropy}~\cite{Astropy2022}.
    }
    \label{fig:peanut_sketch}
\end{figure}

The LHAASO's Peanut source is a~giant, asymmetrically extended gamma-ray emission region at a~Galactic latitude of $b\approx -17.5^\circ$~\cite{Cao2026}. Gamma-ray excess significances are $9.7\sigma$ and $16.0\sigma$ in the $25 < \epsilon < 100~\TeV$ and \mbox{$\epsilon>100~\TeV$} energy ranges, respectively. Peanut's morphology is characterized by a~strip approximately aligned with the Galactic plane, with angular dimensions of $0.45^\circ \times 4.6^\circ$ and two hot~spots, one at the western\footnote{In the following text, we refer to West and East in the Galactic coordinate frame if not stated otherwise.} edge and the other near the center. We show the LHAASO's Peanut\footnote{LHAASO public data: \href{https://english.ihep.cas.cn/lhaaso/pdl}{english.ihep.cas.cn/lhaaso/pdl}} in the Galactic coordinates in~\autoref{fig:peanut_sketch}. A~cross-match of the LHAASO observations with GeV- and TeV-energy gamma-ray source catalogs and pulsar catalogs identified PSR~J0218+4232 as the sole potential accelerator capable of supplying the required power~\cite{Cao2026}.

\emph{PSR J0218+4232} (hereafter J0218) is a~radio luminous millisecond pulsar located at Galactic coordinates $l=139.5^\circ$ and $b=-17.5^\circ$~\cite{Navarro1995} with period $P\simeq 2.3\dm{ms}$ and spin-down luminosity $\dot{\mathcal{E}}\simeq2.4\times10^{35}\dm{erg~s^{-1}}$.
The distance to the pulsar has been estimated using several independent techniques, including the dispersion measure~(DM)~\cite{Navarro1995,Bassa2003,Yao2017}, observations of its white-dwarf (WD) companion~\cite{Bassa2003}, and VLBI parallax measurements~\cite{Du2014} corrected for the Lutz-Kelker bias~\cite{Verbiest2014,Igoshev2016}. The most recent of these estimates are mutually consistent and agree with the distance predicted by recent Galactic electron-density models~\cite{Yao2017} (see Appendix~\ref{appendix:pulsar} for a~more detailed discussion). \emph{Throughout this work, we adopt $d=3\dm{kpc}$.} This distance places J0218 in the Galactic halo, $0.9\dm{kpc}$ away from the Galactic plane; the measured proper motion~\cite{Du2014} corresponds to a~projected peculiar velocity of approximately $100\dm{km~s^{-1}}$ nearly parallel to the Galactic plane and directed approximately westward, away from the Peanut. 

J0218 is one of the most luminous pulsars of the Third \textit{Fermi} Large Area Telescope Pulsar Catalog\footnote{3PC: \href{https://fermi.gsfc.nasa.gov/ssc/data/access/lat/3rd_PSR_catalog/}{fermi.gsfc.nasa.gov/ssc/data/access/lat/3rd\_PSR\_catalog/}} (3PC)~\cite{Smith2023a}. The integrated energy flux between $0.1$ and $100~\GeV$ is $F_\rm{ LAT}=(48.0\pm1.2)\times10^{-12}\dm{erg~cm^{-2}~s^{-1}}$, corresponding to an isotropic gamma-ray luminosity \mbox{$L_\rm{LAT} \approx 0.52\times10^{35}\dm{erg~s^{-1}}
d_3^2$}, where $d_3 = d / (3\dm{kpc})$. This comprises approximately $20\%$ of the pulsar spin-down luminosity. The MAGIC collaboration searched for high-energy gamma-ray signal above $100~\GeV$ from the pulsar itself~\cite{Acciari2021}. No evidence of emission was detected, yielding upper limits extending to~$\sim30~\TeV$.

The Peanut source is seen at much higher energies: its spectrum starts at around $20~\TeV$ and extends beyond $700~\TeV$. The highest-energy photon was detected near the center of the strip, with an energy of $760_{-40}^{+60}~\TeV$ (see~\autoref{fig:peanut_sketch}). Within the spatial boundaries of the Peanut, LHAASO reported 14~photons above $400~\TeV$, compared with 3.2 expected background events. A~recent 40~h VERITAS observing campaign, combined with 140~h of archival observations of the J0218, yielded no detection of the gamma-ray filament and established upper limits below $10~\TeV$~\cite{Wong2026}.

At the distance of 3~kpc, the Peanut has projected length of $\ell_P = 240~\pc$, its first hot spot lies at a~projected distance of $\ell_{P1}\sim20~\pc$, which is slightly lower in $25 < E < 100~\TeV$ bin and slightly higher above $100~\TeV$. The second hot spot is located below the line $b=-17.5^\circ$ at the projected distance of $\ell_{P2} \sim120~\pc$. The Peanut is thus one of the most extended gamma-ray structures seen in the sky, in the same list with the gamma-ray nebula of V4641~Sgr~\cite{Alfaro:2024cjd,LHAASO:2024psv,HESS:2025uyo}, the Cygnus-X region~\cite{Cygnus2024}, and the Galactic center~\cite{Aharonian2006}.

The total gamma-ray luminosity of the Peanut above 20 TeV is estimated to be $L_\rm{\gamma,>20\TeV}\approx(8\times10^{32}\dm{erg~s^{-1}})~d_3^{2}$, while the electron population energy above $50~\TeV$ required to emit the observed $\gamma$-rays via IC scattering on CMB has to be $W_e\sim 10^{45}\dm{erg}~d_3^2$~\cite{Cao2026}. This energy corresponds to approximately 150 years of the pulsar's current spin-down power. The fraction of spin-down power that must be converted into high-energy electrons therefore depends on the particle transport and cooling timescales, which we examine in the following sections.

A~pulsar with spin-down luminosity $\dot{\mathcal{E}}$ injects a~relativistic wind, which, being confined by the surrounding interstellar medium (ISM), forms a~termination shock (TS) and a~corresponding pulsar wind nebula (PWN)~\cite{Bykov2017}.
With YMW16 electron density model\footnote{YMW16 calculator: \href{https://www.atnf.csiro.au/research/pulsar/ymw16/}{atnf.csiro.au/research/pulsar/ymw16/}}~\cite{Yao2017}, we evaluate the ISM density assuming local charge neutrality to be
\mbox{$\rho_\rm{ISM} \simeq 10^{-2}m_p\dm{cm^{-3}}$ at $3\dm{kpc}$}.
The TS stand-off radius is then estimated as
\begin{subequations}
\label{eq:standoff}
    \begin{align}
        R \simeq \sqrt{\dot{\mathcal{E}}/6\pi c\rho_\rm{ISM}v_\rm{rel}^2}
        &= (0.1\dm{pc})~\dot{\mathcal{E}}^{1/2}_{35}\rho^{-1/2}_{-2}v^{-1}_\rm{100},\\
        R_\rm{J0218} &\simeq 0.15~\pc,
    \end{align}
\end{subequations}
where $\dot{\mathcal{E}}_{35} = \dot{\mathcal{E}}/(10^{35}\dm{erg~s^{-1}})$,\\
$\rho_{-2} = \rho_\rm{ISM} / (10^{-2}m_p\dm{cm^{-3}})$,\\
$v_{100} = v/(100\dm{km~s^{-1}})$.

If the PWN is elongated significantly along the pulsar motion, which is directed at a~small angle \mbox{($\sim10^\circ$)} to the Peanut axis, this might explain the shift of the first gamma-ray hot spot similarly to~\cite{Krumholz2024}.

Both the total power of the electron population and the high energy of the detected photons require efficient particle acceleration. In terms of Hillas criterion, by comparing the particle's Larmor radius to the TS radius, we estimate the maximal energy~\cite{Bykov2017}
\begin{subequations}
    \begin{align}
        E_\rm{max,TS} &= e\sqrt{\beta\dot{\mathcal{E}}/\Gamma^2c} 
        = (0.6~\PeV)~\beta^{1/2}\Gamma^{-1}
        \dot{\mathcal{E}}_{35}^{1/2};\\
        E_\rm{max,TS}^\rm{J0218} &\lesssim 1~\PeV,
    \end{align}
\end{subequations}
where $\beta c$ is the flow velocity,\\
and $\Gamma$ is the corresponding Lorentz factor.

These estimates show that acceleration of particles to the observed energies is possible, but requires a~very efficient acceleration mechanism. The discussion in the following sections is dedicated to particle propagation in the regular magnetic field with turbulent component. We remain agnostic to the choice of the acceleration mechanism and assume the high-energy particles accelerated up to $\sim\PeV$ energies following LHAASO's best-fit injected electron spectrum~\cite{Cao2026}.

    \section{Peanut source and the structure of the Galactic Magnetic field}
\label{sec:GMF}

We assume that electrons are accelerated in the vicinity of the pulsar, subsequently captured by the interstellar magnetic field (MF), and transported along its field lines. To investigate this scenario, we must specify our treatment of the Galactic magnetic field (GMF).

We decompose the GMF into a regular large-scale component, $\v{B}_0$, and a turbulent component, $\delta\v{B}$, such that $\v{B}=\v{B}_0+\delta\v{B}$. The field in the vicinity of the Peanut is then characterized by its total strength,
\mbox{$B_\rm{tot}=\sqrt{B_0^2+\lrangle{\delta B^2}}$},
and the turbulent-to-regular ratio,
\mbox{$\eta=\sqrt{\lrangle{\delta B^2}}/B_0$}.
According to the existing GMF models, the MF strength in the Galactic halo is typically at the level of a few $\uG$, while the turbulent and regular components can have comparable amplitudes~\cite{Jansson:2012rt}.

The observed alignment of the Peanut with the Galactic plane motivates us to assume that the regular magnetic field in its vicinity is parallel to the plane or inclined to it at a relatively small angle. At high Galactic latitudes, corresponding to heights $|Z|\gtrsim1\dm{kpc}$ above the plane, the regular field in the outer Galaxy is generally expected to be dominated by the toroidal component. This feature is present in all large-scale GMF models~\cite{1997A&A...322...98H,Han:1999vi,Han:2006ci,Pshirkov:2011um, Jansson:2012pc, Han2024, Unger:2023lob, Korochkin:2024yit}.
If at the location of the Peanut, the regular MF direction is consistent with the large circle denoted with the black dashed line in~\autoref{fig:location}, this source can independently confirm the existence of a toroidal field in this part of the Galactic halo. However, the region is also rather close to the Galactic plane, where the disk becomes increasingly important, despite its strongly model dependent structure. For instance, in the KST24 model~\cite{Korochkin:2024yit}, the disk-field contribution from the Perseus arm dominates at the location of the Peanut, so a gamma-ray emission elongation aligned with the Perseus arm is preferred (shown with a solid black line in~\autoref{fig:location}).

The turbulent field in this region is also poorly constrained by observations. We therefore investigate two idealized models of incompressible turbulence: isotropic Kolmogorov turbulence~\cite{Kolmogorov1941,Giacinti2018}, with an energy spectrum \mbox{$\mathcal{E}(k)\propto k^{-5/3}$}, and anisotropic GS turbulence~\cite{GoldreichSridhar1995, Yan2002, Giacinti2017}, with a perpendicular spectrum \mbox{$\mathcal{E}(k_\perp)\propto k_\perp^{-5/3}$}. Although both models have a Kolmogorov-like inertial-range spectrum, they differ in the spatial distribution of fluctuations. In the isotropic model, the spectrum depends only on the wavenumber magnitude $k$, whereas GS turbulence is anisotropic with respect to the local magnetic-field direction. In the latter, eddies are elongated along the magnetic field, with parallel and perpendicular wavenumbers related by the critical-balance scaling $k_\parallel\propto k_\perp^{2/3}$. These differences affect magnetic-field-line geometry and consequently influence the transport of high-energy electrons~\cite{Giacinti2017}.

Assuming that the Peanut's gamma-ray emission originates from IC scattering of CMB photons by electrons propagating along magnetic field lines (MFLs), we investigate the sensitivity of the field-line geometry to the two aforementioned models. We restrict this geometric study to large-scale turbulent modes by imposing a maximum wavenumber $k_\rm{max}$, corresponding to a minimum wavelength of a few parsecs, well above the typical electron Larmor radius. The adopted $k_\rm{max}$ depends on the turbulence outer scale $L_\rm{max}$. We consider several values of the turbulent-to-regular magnetic-field ratio $\eta$.

To assess the probability of obtaining a structure resembling the Peanut, we exploit the remarkable straightness of the observed gamma-ray filament.
For \mbox{$\eta=0.2,~0.5,~0.7,$ and $1.0$}, we generate 10 magnetic field cubes for each of the two turbulence models (see Appendix~\ref{appendix:mf-generation} for details). In each cube, we compute 500 MFLs starting from randomly selected positions, uniformly distributed in the cubes, yielding 5\,000 MFL realizations for each value of $\eta$ and turbulence model.

\begin{figure}[t!]
    \centering
    \includegraphics[width=\linewidth]{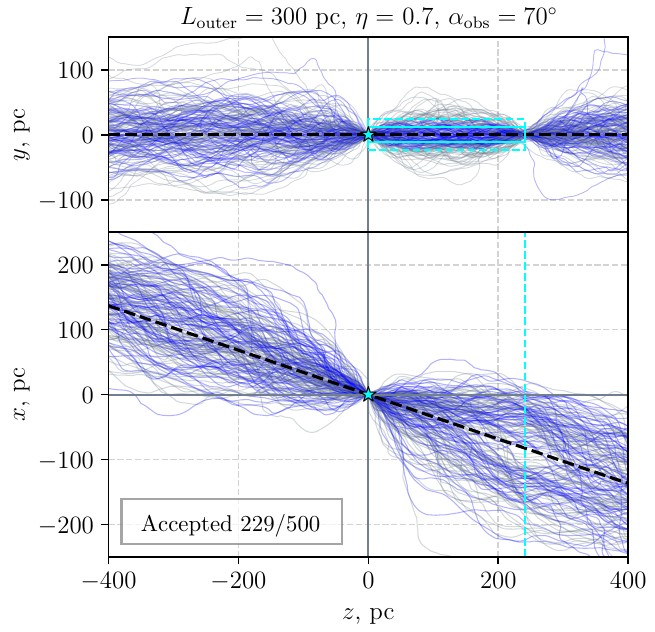}
    \caption{
        Accepted (blue) and rejected (gray) lines for anisotropic turbulent magnetic field (GS) with $L_\rm{max} = 300~\pc$, $\eta = 0.7$, $\alpha_\rm{obs} = 70^\circ$ within a single generated MF cube. A solid cyan rectangle corresponds to the LHAASO strip $240\times24~\pc$,  a dashed cyan rectangle is the extended LHAASO strip $240\times48~\pc$. A cyan star at $(0,0)$ denotes the position of~the~J0218.
        }
    \label{fig:accepted_example}
\end{figure}

These realizations are viewed from different observer angles $\alpha_\rm{obs}$.
To account for the unknown orientation of the regular magnetic field, we rotate each MFL about the regular-field direction so that its projected endpoint at $\ell_P=240~\pc$ lies on the $z$-axis. This procedure is equivalent to varying the orientation of the regular magnetic field relative to the observer. In~\autoref{fig:accepted_example}, we show front and top views of a set of MFLs. We define the $x$-axis along the line of sight (LOS) and the $z$-axis perpendicular to it and lying in the Galactic plane. We accept an MFL if its projection remains within a rectangle of dimensions $240~\pc\times48~\pc$, corresponding to the LHAASO rectangular strip extended to twice its observed width. This extension accounts approximately for the instrumental point-spread function (PSF; see~\autoref{tab:peanut_with_j0218}).

\begin{figure}[t!]
    \includegraphics[width=\linewidth]{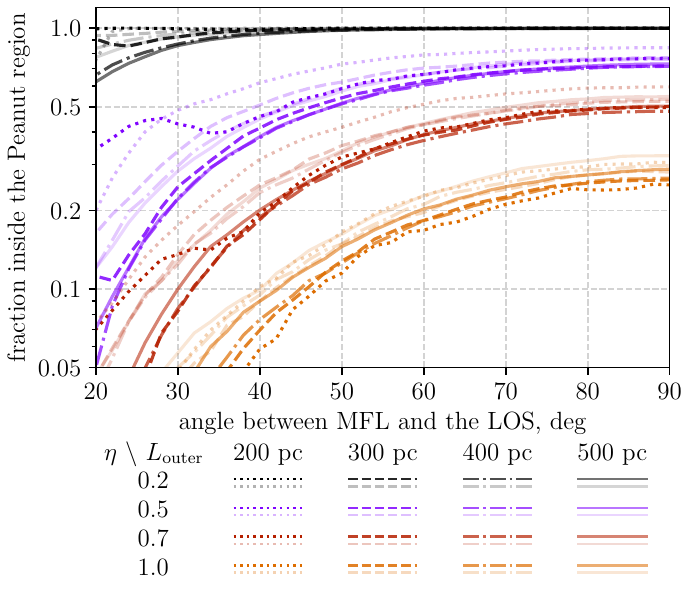}
    \caption{
        Fraction of MFLs inside the extended Peanut rectangle for different values of $L_\rm{max}$, $\eta$, and $\alpha_\rm{obs}$. GS turbulence lines are shown with opaque lines, semitransparent ones correspond to Kolmogorov turbulence.
        }
    \label{fig:line_acceptance_all}
\end{figure}

\autoref{fig:line_acceptance_all} shows the fraction of MFLs that remain within the extended LHAASO strip as a function of the observation angle $\alpha_\rm{obs}$, for different values of $L_\rm{max}$ and $\eta$. Most MFLs deviate from the strip due to magnetic-field fluctuations on smaller scales.

The fraction of accepted lines increases when
\begin{enumerate}
    \item the turbulent-to-regular field ratio $\eta$ decreases, making the field lines more closely follow the regular field;
    \item the observation angle approaches $90^\circ$, such that the relevant projected scales correspond to smaller turbulent wavelengths, where the fluctuation power is lower.
\end{enumerate}

In addition, \autoref{fig:line_acceptance_all} shows no significant differences between the two turbulence models: Kolmogorov and GS turbulence yield similar accepted fractions across the considered values of $\eta$ and $L_\rm{max}$.

Our simulations, described in the following sections, indicate that for Kolmogorov and GS scenarios the origin of the Peanut hot spots might be connected to magnetic bottling and geometric effects, respectively. In both cases, the substantial flux difference between the hot spot regions and the strip makes us apriori prefer higher values of $\eta$. At the same time, even corrected by the magnetic field direction, which we free, the turbulent modes of scale lower than the scales of the Peanut become large enough to direct the line outside of the LHAASO srip. As seen from~\autoref{fig:line_acceptance_all}, the accepted fraction of lines drops from effectively 1 at $\eta=0.2$ down below 0.3 for $\eta = 1.0$.

This straight line fraction has to be taken into account when considering the overall probability to reproduce the main Peanut features. In the two following sections we present different approaches to account for these geometric effects: in the diffusive case, the observer direction is optimized, which allows for observation of the Peanut in the plane, where its perpendicular variations are the smallest; in the ballistic scenario, we first fix the geometry of the MFL to the Peanut end and its second hot spot, and optimize the predicted signal after.

    \section{Diffusion\\in Kolmogorov turbulence}
\label{sec:diffusion}

In this section, we assume that the turbulent component of the GMF is 3D isotropic Kolmogorov turbulence. For turbulent-to-regular ratios of the order of 1, the Peanut length exceeds the mean free path, hence the diffusion approximation is applicable. The elongation of the Peanut suggests strongly anisotropic diffusion, which is realized in scenarios of the magnetic field with regular and turbulent components~\cite{Giacinti2018}. We choose initial axes in such a way that the $X$-axis is along the regular field.

Normally, in the anisotropic diffusion scenario, one expects to see two-sided emission around the source~\cite{2013GG}. To provide the conditions necessary for observation of a~clearly one-sided emission, we assume that the injection of high-energy particles at the pulsar is one-sided but isotropic at $X>0$. In addition, we assume the existence of an~ideal magnetic mirror at $X=0$, blocking particles from going on the other side of the MFL.

In the following subsections, we calculate timescales in the diffusion scenario, derive timescale-driven analytical constraints, describe our approach to complete particle propagation simulation in the Kolmogorov turbulence and compare simulations with the LHAASO observations.

For compactness, in the following text we use some specific notation, summarized in~\autoref{tab:notation}.

\begin{table}[h!]
    \centering
    \begin{tabular}{c|c|c}
        Symbol & Transcription & Quantity \\
        \hline
        $S_\alpha$ & $\sin\alpha$ &
        \multirow{3}{*}{Functions of the pitch-angle $\alpha$} \\
        $C_\alpha$ & $\cos\alpha$ & \\
        $T_\alpha$ & $\tan\alpha$ & \\
        \hline

        $d_{3}$  & $d / (3\dm{kpc})$ &
        Distance to the pulsar \\

        \hline

        $E_{15}$ & $E / (1~\PeV) $ &
        Electron energy \\
        
        $B_{-6}$ & $B / (1~\uG) $ & 
        Magnetic field strength \\

        $D_{30}$ & $D_\prl / (10^{30}\dm{cm^{2}~s^{-1}}$) &
        Parallel diffusion coefficient \\
        
        $L_{100}$ & $L_\rm{max} / (100~\pc)$ & 
        Turbulence outer scale \\

        \hline
        $W_{e,45}$ & $W_e / (10^{45}\dm{erg})$ & Electron population energy \\

        $\dot{\mathcal{E}}_{35}$ & $\dot{\mathcal{E}} / (10^{35}\dm{erg ~s^{-1}})$ & Pulsar spin-down luminosity 
        
    \end{tabular}
    \caption{Notation transcription table.}
    \label{tab:notation}
\end{table}

\subsection{Timescales in the diffusion scenario}

To derive the timescales and analytical constraints, we assume the MFL to be a~straight line, MF strength is considered constant along the line. The angle between the LOS and the MFL is denoted as $\alpha_\rm{obs}$ similar to~\autoref{fig:accepted_example}.

\entry{Cooling timescales}

The highest-energy gamma-ray from the Peanut reported by LHAASO has the energy of $760_{-40}^{+60}~\TeV$~\cite{Cao2026}. At this energy, IC scattering is already deep in Klein-Nishina regime, hence the upscattered photon energy is close to the incident electron energy, while the electron's lifetime is defined primarily by synchrotron losses. For a~$760~\TeV$ electron diffusing in magnetic field $B$, the cooling timescale is given by~\cite{BlumenthalGould1970}
\begin{subequations}
    \begin{align}
        \tau_\rm{syn} = \dfrac{9(mc^2)^4}{4e^4c}\dfrac{1}{E B^2}
        & \approx  12.5~\rm{kyr}~B_{-6}^{-2}E_{15}^{-1} \\
        \tau_\rm{syn,760\TeV} &\approx 16.5\dm{kyr}~B_{-6}^{-2}.
    \end{align}
\end{subequations}

The LHAASO significance maps show that the Peanut extends up to $4.5^\circ$ eastward both above and below $100~\TeV$ (see~\autoref{fig:peanut_sketch}). For a~conservative estimate, we assume that $100~\TeV$ electrons reach the end of the nebula and compute their cooling timescales. At this energy, IC cooling on CMB is non-negligible and can be treated as synchrotron losses in an effective isotropic magnetic field of
\mbox{$B_\rm{CMB,100\TeV} = (\dot{\mathcal{E}}_\rm{IC}/\dot{\mathcal{E}}_\rm{syn,1\uG})^{1/2}~\uG\simeq 1.86~\uG $}.
Effective magnetic field $B_\rm{ef}$ then comprises both actual and IC-related fields: $B_\rm{ef} = (B^2 + B_\rm{CMB}^2)^{1/2}$.

\begin{subequations}
    \begin{align}
        \tau_\rm{syn+IC} = \dfrac{9(mc^2)^4}{4e^4c}
        \dfrac{1}{E B_\rm{ef}^{2}}
        & \approx  12.5~\rm{kyr}~B_\rm{ef,-6}^{-2}E_{15}^{-1} \\
        \tau_\rm{syn+IC,100\TeV} &\approx 36\dm{kyr}
        \pbr{1 + \frac{B_{-6}^2}{3.5}}^{-1}
    \end{align}
    \label{eq:IC_syn_100TeV}
    \vspace{-3ex}
\end{subequations}

\entry{Diffusion timescales}

In the diffusion scenario, length and width of the Peanut are determined by diffusion of particle trajectories. Kolmogorov turbulence predicts~\cite{Giacinti2018} diffusion coefficient scaling to be
\begin{equation}
    D_\prl = D_{\prl,1\PeV} L_{100}^{2/3} \eta^{-2} E_{15}^{1/3} B_{-6}^{-1/3}.
    \label{eq:parallel_diffusion_etaouter}
\end{equation}
where $D_{\prl,1\PeV}$ is the parallel diffusion coefficient at $1~\PeV$ and has typical values of $D_{\PeV} \simeq 10^{30}\dm{cm^2~s^{-1}} \approx 3.3\dm{pc^2~yr^{-1}}$ for magnetic field of $\eta = 1.0$~\cite{Giacinti2018,Mertsch2020}. 

As already mentioned, LHAASO observes photons below and above $100~\TeV$ at the projected distance $\ell_{P,100\TeV}\approx240~\pc~d_3$. At the same time, the highest-energy gamma-ray was observed near the second hot spot of the Peanut at the projected distance \mbox{$\ell_{P,760\TeV} = 100~\pc~d_3$}. Diffusion timescales $\tau_D = \ell^2 / 4D$ for the corresponding electrons are
\begin{subequations}
    \begin{align}
        \tau_{D,100\TeV} &= 9.36\dm{kyr}~D_{30}^{-1} d_3^2\Sa^{-2}B_{-6}^{1/3} ; \\
        \tau_{D,760\TeV} &= 0.83\dm{kyr}~D_{30}^{-1} d_3^2\Sa^{-2}B_{-6}^{1/3} .
    \end{align}
\end{subequations}

\entry{Transport timescales}

Diffusion is considered steady after the particle experienced several scatterings, whose frequency is characterized by timescale $\tau_\prl$. As the diffusion coefficient is defined as $D_\prl \simeq c\lambda_\prl/3 = c^2\tau_\prl/3$, these timescales are
\begin{subequations}
    \begin{align}
        \tau_\prl &= 0.11\dm{kyr}~D_{30} B_{-6}^{-1/3} E_{15}^{1/3};\\
        \tau_{\prl,760\TeV} &= 0.10\dm{kyr}~D_{30} B_{-6}^{-1/3}; \\
        \tau_{\prl,100\TeV} &= 0.05\dm{kyr}~D_{30} B_{-6}^{-1/3}.
    \end{align}
\end{subequations}

\entry{Power timescale}

The fraction $\xi$ of the pulsar's spin-down luminosity $\dot{\mathcal{E}}$ being pumped into high-energy electrons above $50\dm{TeV}$ has a~timescale
\begin{subequations}
    \begin{align}
        \tau_\rm{pow} = W_e / f_e \xi\dot{\mathcal{E}} &\approx 0.32\dm{kyr}~W_{e,45}\,d_3^2\,\dot{\mathcal{E}}_{35}^{-1}\xi^{-1} f_e^{-1}; \\
        \tau_\rm{pow}^\rm{J0218} &\approx 0.13\dm{kyr}~d_3^2\,\xi^{-1} f_e^{-1},
    \end{align}
\end{subequations}
where $f_e$ is the fraction of electrons accumulated within the Peanut region.

\subsection{Analytical constraints in the diffusion scenario}

We choose to express the following constraints in terms of the parallel diffusion coefficient normalization $D_{30}$.

\entry{Cooling vs. diffusion}

For energetic particles to reach their furthest elongation points, we require $\tau_\rm{syn,760\TeV} > \tau_{D,760\TeV}$. This gives
\begin{equation}
    D_{30} > 0.05~d_3^2\Sa^{-2}B_{-6}^{7/3}.
    \label{eq:diffusion_760_constraint}
\end{equation}

By demanding $\tau_\rm{syn+IC,100\TeV} > \tau_{D,100\TeV}$, we obtain
\begin{equation}
    D_{30} > 0.26~d_3^2\Sa^{-2}B_{-6}^{1/3}(1 + B_{-6}^2 / 3.5).
    \label{eq:diffusion_100_constraint}
\end{equation}

We find that constraint~\eqref{eq:diffusion_100_constraint} dominates over~\eqref{eq:diffusion_760_constraint}. This can be motivated by the fact that the minimal diffusion coefficient obtained by equating loss and diffusion times scales roughly as $\ell_P^2 E^{2/3}$, and in timescale comparison, distance-ratio factor $2.4^2 =5.76$ dominates over energy-ratio factor $7.6^{2/3} \approx 3.87$. Hence, in the diffusion scenario, the remoteness of lower energy particles places more stringent constraints than the highest-energy observations.

\entry{Diffusion vs. free-propagation}

For the diffusion approximation to be applicable, the diffusion timescale has to exceed the transport timescale: $\tau_D > \tau_\prl$. This constraint is dominated by the $760\dm{TeV}$ photon, which places an upper limit on diffusion coefficient normalization:
\begin{equation}
    D_{30} < 2.9~B_{-6}^{1/3} d_3\Sa^{-1}.
    \label{eq:freeprop-dif}
\end{equation}

\entry{Power constraints}

We assume that the electron population is injected in a steady-state regime at the source at the rate $Q$. For the~soft best-fit LHAASO spectrum with $p\approx2.5$, the electron population energy is dominated by low-energy particles ($E\lesssim100~\TeV$). For a~simple estimation, we assume electrons to disappear after lifetime $\tau = \tau_\rm{syn+IC,100\TeV}$. With the mirror at the source, 1D particle distribution along the MFL $N_e(s)$ is given by
\begin{equation}
    N_e(s) = \frac{4Q\tau}{\lambda}\left[\frac{1}{\sqrt{\pi}} e^{-\frac{s^2}{\lambda^2}} - \frac{s}{\lambda}\rm{erfc}\pbr{\frac{s}{\lambda}}\right],
    \label{eq:1D-diffusion-profile}
\end{equation}
where $\lambda = \sqrt{4D\tau}$.
The fraction $f_e(\ell)$ of particles at $100~\TeV$ in the visible part of the nebula $s < \ell$ is determined by the ratio of diffusion and cooling timescales \mbox{$r = \tau_{D,100\TeV}/\tau_\rm{syn+IC,100\TeV} < 1$}. This fraction can be approximated as $f_e \approx 0.94r^{1/4}$ with $10\%$ accuracy \mbox{for~$r < 2$} (see Appendix~\ref{appendix:diffusion-profile}).

By equating power and diffusion timescales, we estimate the required power fraction $\xi$ transferred to HE particles: $\tau_\rm{pow}^\rm{J0218} = \tau_{D,100\TeV}.$ We obtain
\begin{subequations}
    \begin{equation}
        \xi \approx 0.02~D_{30}^{5/4}d_3^{-1/2}\Sa^{5/2}B_{-6}^{-5/12}(1 + B_{-6}^2/3.5)^{-1/4};
        \label{eq:power-constraints-diffusion-xi}
    \end{equation}
    \vspace{-5ex}
    \begin{equation}
        D_{30} \approx 22~\xi^{4/5} d_3^{2/5} \Sa^{-2} B_{-6}^{1/3}(1 + B_{-6}^2/3.5)^{1/5}.
        \label{eq:power-constraints-diffusion-D}
    \end{equation}
    \label{eq:power-constraints-diffusion}
\end{subequations}

Eq.~\eqref{eq:power-constraints-diffusion-D} can be used as an upper limit for the diffusion coefficient. For $\xi < 0.1$ these constraints are weaker than~\eqref{eq:freeprop-dif}. Thus, we conclude that given aforementioned constraints, the spin-down luminosity fraction for feeding the Peanut lies within the boundaries of $0.4\% < \xi < 10\%$, which implies that particle acceleration has to be rather efficient.

\entry{Diffusion limit summary}

\begin{figure}[h!]
    \centering
    \includegraphics[width=\linewidth]{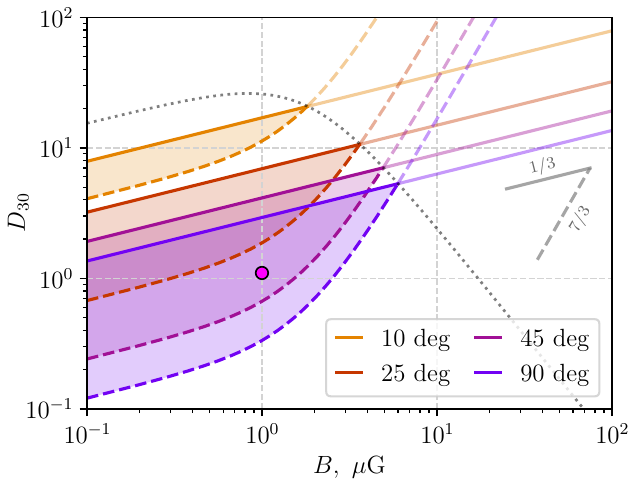}
    \vspace{-5ex}
    \caption{Allowed diffusion coefficient normalization $D_{30}$ regions for different magnetic fields at several angles $<90^\circ$.  The dotted line shows maximal allowed diffusion coefficient value for each given magnetic field, obtained by equating~\eqref{eq:diffusion_100_constraint} with~\eqref{eq:freeprop-dif} and subsequent elimination of the observation angle parameter. A~pink circle at $(1.0\uG, 1.1)$ shows the parameters we investigate in the simulation.
    }
    \label{fig:diffusion_angles_analytic}
\end{figure}

\begin{figure*}[!t]
    \centering
    \includegraphics[width=\linewidth]{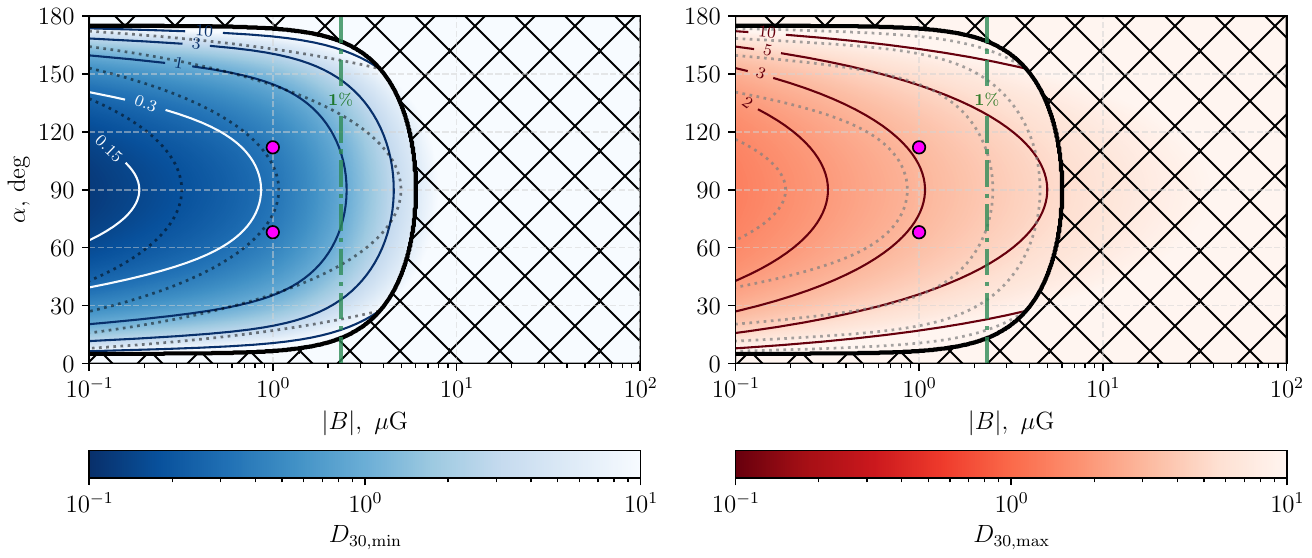}
    \vspace{-3ex}
    \caption{
        Analytical constraints on minimal and maximal diffusion coefficient normalization in the diffusion regime in observed angle - magnetic field strength parameter space. In both panels, the region where $D_\rm{30,min} > D_\rm{30,max}$ is hatched in black. Vertical dashed line indicates the MF strength at which the minimal power can be less than $1\%$. Two~pink circles at $(1.0\uG, 68^\circ)$ and $(1.0\uG, 112^\circ)$ show the parameters of our best-fit simulation. \\
        \emph{Left:} Minimal diffusion coefficient~\eqref{eq:diffusion_100_constraint} (shades of blue).
        \emph{Right:} Maximal diffusion coefficient~\eqref{eq:freeprop-dif} (shades of red).
    }
    \label{fig:diffusion_analytic}
    \vspace{-2ex}
\end{figure*}

The derived constraints are summarized in~\autoref{fig:diffusion_angles_analytic} and~\autoref{fig:diffusion_analytic}. We see that the diffusion regime can exist within rather stringent bounds of magnetic fields $B < 7~\uG$ and parallel diffusion coefficients at $1~\PeV$ and $1~\uG$ between $10^{29}$ and $10^{31}\dm{cm^2~s^{-1}}$, which constrains selection of possible values for $\eta$ and $L_\rm{max}$ via eq.~\eqref{eq:parallel_diffusion_etaouter}. When the actual diffusion coefficient is close to minimal $D_\rm{min}$ given by~Eq.~\eqref{eq:diffusion_100_constraint} (dashed lines in~\autoref{fig:diffusion_angles_analytic} and left panel in~\autoref{fig:diffusion_analytic}), the Peanut length is constrained by cooling, and one does not expect to see any prolongation of the source with more data. For larger diffusion coefficients, the Peanut length may be constrained by MFL geometry.

From the energetic perspective, the diffusion regime requires efficient particle acceleration by the Pulsar, as the electron injection power is constrained between $0.4\%$ and $10\%$ of the pulsar's spin-down luminosity. In~\autoref{fig:diffusion_analytic}, we show the boundary magnetic field of $B_{1\%} \approx 2.3~\uG$, above which for any available diffusion coefficient, injected electron energy fraction has to exceed $1\%$.

\subsection{Diffusion regime simulations}

To demonstrate that the Peanut source can be reconstructed in the Kolmogorov scenario, we model propagation of high-energy electrons by computing their individual trajectories in three-dimensional realizations of turbulent magnetic fields.

The turbulent MFs are generated on a~set of nested Cartesian grids following the method described in Refs.~\cite{2012GG,2013GG}. This approach allows a~wide range of spatial scales to be included in the turbulent field.  In addition to the turbulent component, we include a~constant regular magnetic field. In the Kolmogorov simulations used here, the regular component is added along the $X$-axis of the simulation frame. We adopt $B_{\rm {tot}}=1~\mu{\rm G}$, $\eta=1.0$, and outer scale of $L_\rm{max} = 500~\pc$. Contrary to the MF generation \mbox{in~\autoref{sec:GMF}}, where we were interested in the field geometry at Peanut scales, for particle propagation, the turbulent magnetic field is constructed using two nested grids~($256^3$), yielding a~finest grid spacing of $L_{\min}\simeq 0.0076~\pc$, smaller than the characteristic Larmor radius of injected electrons in a~MF of~$1~\uG$.

As mentioned before, we assume one-sided and isotropic injection along the $X$-axis; for each MF configuration, we inject 5\,000 particles at the position of J0218. The injected electron spectrum is chosen to be the same as that adopted in Ref.~\cite{Cao2026}.

Instead of solving the diffusion-loss equation, we compute individual trajectories of high-energy electrons. This test-particle approach follows electron motion directly in prescribed magnetic-field configurations, allowing us to connect the resulting gamma-ray morphology to the physical parameters of the turbulence. Synchrotron and IC losses are taken into account, within the continuous energy-loss approximation. To mimic continuous injection from the source, we save electron locations and momenta (i.e. their energies) at regular time intervals, following~Ref.~\cite{Lopez-Coto2018}.

For each simulation, we compute projected gamma-ray maps for 192~viewing directions using \texttt{HEALPix}~\cite{healpix} with~$N_\rm{side}=4$. For a~given viewing direction, the simulated particle positions are projected onto the plane of the sky and binned into two-dimensional RA-Dec maps. For each recorded particle point, the electron energy is calculated from its momentum, and the corresponding IC gamma-ray weight is computed in the $25 < \epsilon < 100~\TeV$, $\epsilon >100~\TeV$, and $\epsilon > 400~\TeV$ gamma-ray bands using~\texttt{GAMERA}~\cite{GAMERA} under assumption of isotropicity of the local electron distribution. The simulated maps are then smoothed with Gaussian kernels to mimic the angular resolutions of WCDA and KM2A in the corresponding energy bands. We then compare the resulting morphology with the LHAASO significance contours and the observed right-ascension counts profiles.

In total, we examined 68 Kolmogorov magnetic-field configurations with the same parameters of~\mbox{$L_\rm{min} = 500~\pc$}, $\eta=1.0$, $B_\rm{tot} = 1~\uG$. The best-fit configuration is discussed in the following subsection.

\begin{figure*}
    \centering
    \includegraphics[width=\linewidth]{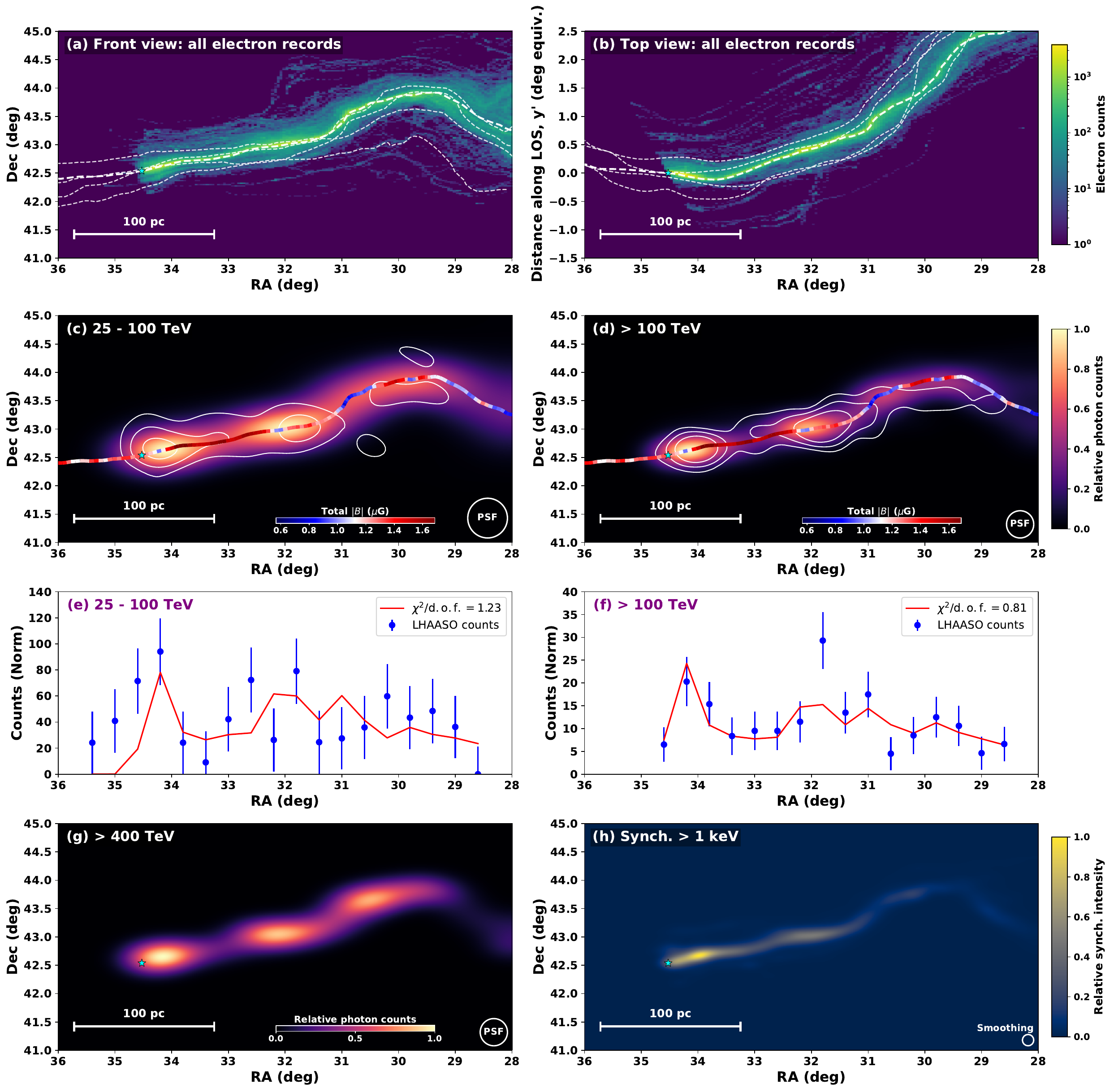}
    \caption{
        Representative best-fit projections from the diffusion regime simulations. The projections correspond to HEALPix with $N_{\rm {side}}=4$ pixel index $85$. The LOS unit vector in the original simulation frame is $\hat{\v{n}}_{\rm{LOS}}=(0.377331,-0.910958,0.166667)$, which corresponds to~$\alpha_\rm{obs}\approx112^\circ$.\\
        \textbf{(a)-(b)}: All electron counts projected onto the RA-Dec plane (front view) and RA-LOS plane (top view) in logarithmic scale. Thick dashed line corresponds to the principal MFL passing through the source; thinner dashed lines represent other MFLs. The cyan star marks the pulsar position. White scale bars correspond to 100 pc at an assumed distance of 3.15 kpc.\\
        \textbf{(c)-(d)}: Simulated IC gamma-ray intensity maps in the $25<\epsilon<100~\TeV$ and $\epsilon>100~\TeV$ bands, respectively. White contours indicate the LHAASO significance levels of 3, 5, and 7$\sigma$. White circles indicate the adopted PSF angular scales. The field strength along the principal MFL is color-coded as indicated by the inset color bar.\\
        \textbf{(e)-(f)}: Projected 1D profiles compared with the observed LHAASO gamma-ray count profiles (blue points) in different energy band, with the corresponding reduced ($\chi^2/\rm{d.o.f.}$) values indicated, respectively. Assuming the errors to be dominated by background statistical fluctuations, we exclude the points with values below the $2\sigma$ upper-limit.\\
        \textbf{(g)-(h)}: Predicted IC emission above 400 TeV and the predicted synchrotron emission above 1~keV, calculated using the local magnetic field component perpendicular to the LOS.
    }
    \label{fig:diffusive_bestfit}
\end{figure*}

\subsection{The best-fit configuration}

Panels~(a) and~(b) of \autoref{fig:diffusive_bestfit} show the simulated electron distributions projected onto the plane of the sky and the plane containing the LOS, respectively. In both projections, the electrons exhibit a highly anisotropic, filamentary distribution that approximately follows the \emph{principal MFL}, which passes through J0218. Shown in the logarithmic scale, the figures reveal a drastic decrease of electron counts in the direction orthogonal to the line. This demonstrates that perpendicular diffusion of electrons from the principal MFL exists, but has a negligible effect on the observed gamma-ray emission.

The gamma-ray maps, smoothed with the LHAASO PSF in each energy bin, are shown in panels~(c) and~(d) of the~\autoref{fig:diffusive_bestfit}. By comparing them with the LHAASO significance contours, we see that our simulations reproduce the two Peanut hot spots. From panel~(b) we conclude that there is no obvious local bending of the MFL at their positions. When we examine the variation of the total MF strength along the principal MFL, we find that both hot spots appear to coincide with regions of relatively weak magnetic field strength, bounded by stronger field regions. Such configurations resemble magnetic bottles, where charged particles with sufficiently large pitch angles undergo magnetic mirroring and experience enhanced local residence times. Such magnetic trapping in turbulent fields produces localized particle overdensities in the distribution of escaping particles. This suggests that magnetic-bottle-like confinement contributes to the formation of the observed gamma-ray hot spots, even in the absence of any pronounced MFL bending. It is interesting to point out that, in our simulations, particles do not remain permanently trapped in those regions, thanks to pitch-angle scattering and cross-field diffusion allowing them to escape eventually.

At higher photon energies, the predicted IC emission above $400~\rm{TeV}$ exhibits three distinct intensity enhancements along the extended structure (panel~(g) of~\autoref{fig:diffusive_bestfit}). The first two are located near the two LHAASO hot spots, while the third appears farther away from the pulsar, coinciding with the MF minimum. This additional hot spot may be qualitatively consistent with the spatial distribution of several-hundred-TeV photons reported by LHAASO~\cite{Cao2026}, although we suggest it to be a property of this particular MF configuration, which necessarily differs from the one realized at the location of the Peanut.

In addition, we calculate the synchrotron emission from the simulated electrons following~\cite{BlumenthalGould1970}, assuming a locally isotropic distribution of electron momentum directions. The local magnetic field vector at each particle position is reconstructed from the nested turbulence grids, including the regular MF component. Due to isotropicity, the effective MF for the synchrotron emission is defined by the component of the MF perpendicular to the LOS.

The resulting synchrotron intensity map is shown in panel~(h) of~\autoref{fig:diffusive_bestfit}. For a better visibility, we smooth it with the Gaussian kernel of size $0.1^\circ$. The predicted emission follows the curved filamentary structure of the propagated electrons, with the brightest region close to the pulsar and weaker emission extending farther along the filament. This spatial variation arises from the combined effects of the high-energy electron distribution, which peaks in the lower-MF regions, and the local MF component perpendicular to the LOS. The two hot spots remain clearly visible on the filament, while their positions appear slightly shifted from the MF minima towards the area where the MF strength increases.


These results demonstrate that electron propagation in Kolmogorov turbulence can produce an extended filamentary morphology of the electron distribution, which allows for reproducing morphology of the Peanut. Other simulations show that without a mirror at $X=0$, the source is necessarily two-sided. The first and second hot spots likely originate from magnetic-bottle-like confinement of electrons in regions with reduced magnetic field line strength. This explanation favors non-negligible turbulent-to-regular magnetic field strengths in the Peanut region. This is in line with the fact that we find it substantially more difficult to reproduce the Peanut's hot spots with Kolmogorov simulations with $\eta \ll 1$.

Gamma-ray hot spots could, in principle, also be induced by MFL bending towards the LOS, which may create hot spots similar to those observed in the Peanut, when the electron density along the line-of-sight is sufficiently large in those regions. This effect nonetheless does not contribute to the formation of the two hot spots in the selected magnetic field configuration shown in~\autoref{fig:diffusive_bestfit}. The relative frequency of these two possible hot-spot creation scenarios and the detectability of the predicted x-ray emission can be studied in the future, through an extensive scan of the parameter space.

    \section{Ballistic propagation\\in Goldreich-Sridhar turbulence}
\label{sec:ballistic}

In this section, we assume the GMF to be a~combination of a~regular field and a~turbulent component described by Goldreich-Sridhar~(GS) turbulence~\cite{GoldreichSridhar1995}. Due to its anisotropic spectrum, resonant scatterings are suppressed in comparison to the Kolmogorov turbulence, which makes timescales to reach diffusion longer than the time particles spend inside the Peanut region~\cite{Giacinti2017}. Thus, in the GS scenario, particle motion within the boundaries of the Peanut can be treated as ballistic.

As previously, for analytical constraints in the diffusion scenario, we assume magnetic field line to be straight and to form an angle of $\alpha_\rm{obs} \in (0,180^\circ)$ to the LOS (similar to~\autoref{fig:accepted_example}). Particle injection at the source is considered to be isotropic in $4\pi$. In the ultrarelativistic limit (UrL), gamma-rays in the IC scattering are emitted along electron's trajectory, hence a terrestrial observer receives photons only from electrons having pitch-angle equal to the observation angle ($\alpha=\alpha_\rm{obs}$). This means that for $\alpha \neq 90^\circ$ the radiation is naturally one-sided, and the observation angles are constrained to be acute, $\alpha_\rm{obs} < 90^\circ$. Contrary to the diffusion scenario, no magnetic mirror close to the source is needed to obtain one-sided emission.

\subsection{Timescales in the ballistic scenario}
\label{subsec:ballistic_timescales}

Below, we list the timescales relevant for the description of the Peanut in the ballistic scenario.

\entry{Cooling timescales}

Due to pitch-angle dependence of the gamma-ray emission, cooling timescales differ from the diffusion regime. As previously, for a~$760~\TeV$ electron, we account for synchrotron losses only:
\begin{subequations}
    \begin{align}
        \tau_\rm{syn} = \dfrac{3(mc^2)^4}{2e^4c}\dfrac{1}{E B^2\Sa^2} &\approx 8.4\dm{kyr}~B_{-6}^{-2}\Sa^{-2}E_{15}^{-1};\\
        \tau_\rm{syn,760\TeV} &\approx 11\dm{kyr}~B_{-6}^{-2} \Sa^{-2}.
    \end{align}
\end{subequations}

For a~100~TeV electron, analogous to eq.~\eqref{eq:IC_syn_100TeV}, we use the notion of effective magnetic field, which has no pitch-angle-averaging factor of~$\sqrt{2/3}$:
\begin{equation}
    \tau_\rm{syn+IC,100\TeV} \approx 36\dm{kyr}\pbr{1 + \frac{B_{-6}^{2}\Sa^2}{2.3}}^{-1}
\end{equation}

\entry{Propagation timescales} 

Ballistic propagation timescale is defined as
\begin{subequations}
    \begin{align}
        \tau_\rm{bal} = \ell_\prl/v_\prl &= \ell_P/c\,\Sa\Ca = 2\ell_P/c\,S_{2\alpha}\\
        \tau_\rm{bal,760\TeV} &\approx 0.66\dm{kyr}~d_3\,S_{2\alpha}^{-1},\\
        \tau_\rm{bal,100\TeV} &\approx 1.6\dm{kyr}~d_3\,S_{2\alpha}^{-1}.
    \end{align}
\end{subequations}

Propagation timescale is the shortest for $\alpha = 45^\circ$.

\entry{Power timescale}

Similarly to the diffusion regime, the timescales are given by eq.~\eqref{eq:power-constraints-diffusion}.

\subsection{Analytical constraints in ballistic regime}

\begin{figure}[b!]
    \centering
    \includegraphics[width=\linewidth]{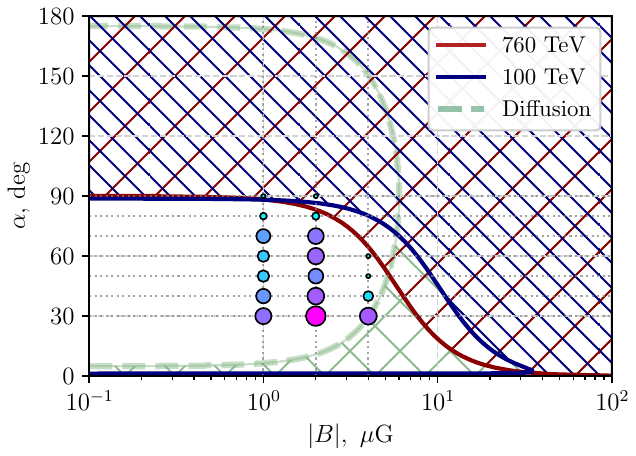}
    \vspace{-3ex}
    \caption{
    Allowed and forbidden (hatched) regions in ballistic scenario derived from $760~\TeV$ photon synchrotron cooling~\eqref{eq:760TeV-ballistic-limit} [red], $100~\TeV$ photon cooling~\eqref{eq:100TeV-ballistic-limit} [blue]. Diffusion limits are shown in pale green for comparison.\\
    Colored circles correspond to the part of the parameter space we investigated with our simulations. The size and the color of the point show relative number of configurations in which we obtain fit quality of~$\chi^2/\rm{d.o.f}. < 1.4$.
    }
    \label{fig:ballistic_limits}
\end{figure}

\entry{Cooling vs. transport timescales}
    
For both 760~TeV and 100~TeV electrons to produce photons at their observed positions, cooling timescales have to exceed ballistic propagation timescales: $\tau_\rm{cool} > \tau_\rm{bal}$. This gives, respectively, 
\begin{subequations}
    \begin{align}
        B &< 5.7~\uG~\Ta^{-1/2}d_3^{-1/2},
        \label{eq:760TeV-ballistic-limit}
        \\
        B &< 10~\uG~\pbr{d_3^{-1}\Ta^{-1}-0.02\Sa^{-2}}^{1/2}
        \label{eq:100TeV-ballistic-limit}
    \end{align}
\end{subequations}

These two equations place constraints on MF strength and observation angles, which are shown in~\autoref{fig:ballistic_limits}. It is interesting to note that contrary to the diffusion case, in the ballistic regime constraints coming from a~$760~\TeV$ photon are stricter than the ones based on observation of $100~\TeV$ photons at the far edge of the Peanut.

\entry{Power vs. propagation timescales}

Contrary to the diffusive scenario, in the ballistic case, particles do not accumulate within the Peanut region, making the power timescale equal to the propagation timescale associated with the length of the region: \mbox{$\tau_\rm{pow} = \tau_\rm{bal,100\TeV}$}, which gives
\begin{equation}
    \xi = 0.08\,S_{2\alpha}.
\end{equation}

Power demand has a~maximum at $\alpha = 45^\circ$, where the propagation is the fastest. At $\alpha$ close to $0^\circ$, the filament has to be very long, and at $\alpha$ close to $90^\circ$ particle propagation along the filament is very slow, both these reasons reduce power demands. We also note that even though this~$<8\%$ fraction of the total spin-down luminosity does not contradict the energetics, it requires a~very efficient acceleration and demands a~substantial fraction of the pulsar power.

\subsection{Particle propagation along magnetic field line
in~the guiding center approximation}
\label{subsec:free-particle-propagation}

To model particle propagation in the ballistic scenario, we use the guiding center theory~\cite{Northrop1963,Petrosian1985,Medvedev2026}. Motion of ultrarelativistic particles in presence of radiative losses is desribed by the following equations (for derivation see Appendix~\ref{appendix:guiding-center}):
\begin{equation}
    \dfrac{ds}{dt} = \mu c;
    \quad
    \dfrac{d\mu}{dt} = -\mathcal{I}_0\dfrac{dB}{ds};
    \quad
    \dfrac{dE}{dt} = -P_\rad,
    \label{eq:final-evolution}
\end{equation}
where $s$ is the coordinate along the line;\\
$\mu = \cos\alpha$ is defined by particle pitch-angle;\\ 
$\mathcal{I}_0 = (1-\mu^2)/2B$ is the adiabatic invariant in the UrL;\\
and $P_\rad$ comprises IC and synchrotron losses.

The first two equations in~\eqref{eq:final-evolution} describe spatial propagation and are independent of energy. This allows for separation of spatial and energy evolution, which we profit from to accelerate the computations (see~Appendix~\ref{appendix:simplified_transport}).

To account for particle cooling, in eq.~\eqref{eq:effective_times_and_fields} we introduce effective propagation time $t_\rm{ef}$ and effective MF strength~$B_\rm{ef}$, which allow for the use of precomputed energy evolution tables (see~Appendix~\ref{appendix:effective-cooling} for more details).
\begin{equation}
    t_\rm{ef} = \intop_\rm{0\to s}\dfrac{d\tilde{s}}{\mu c};
    \quad
    B_\rm{ef}^2 = \intop_\rm{0\to s}\dfrac{d\tilde{s}}{\mu c}B^2(1-\mu^2).
    \label{eq:effective_times_and_fields}
\end{equation}

Electron propagation and computation of IC and synchrotron emission for the given line follows the algorithm below.

\begin{enumerate}[label={\!\arabic*)}]
    \item We select a~precomputed MFL and the observation angle $\alpha_\rm{obs}$. For each point at the selected MFL, we compute the observed angle $\mu_\rm{obs} = \cos\phi_\rm{obs}$, which differs from $\alpha_\rm{obs}$ due to turbulent field fluctuations. We also determine local magnetic field strength $B$, and calculate the invariant $\mathcal{I}_0$. With $\mathcal{I}_0$ at each point and the MF strength $B_0$ at the source, we derive the absolute value of the injection angle \mbox{$|\mu_0| = \sqrt{1 - 2\mathcal{I}_0 B_0}$} for particles observed at each point along the line.

    \item For each point along the MFL, we determine candidate mirror positions, and find possible trajectories starting from the source and ending at the selected point with $\mu_\rm{obs}$. These trajectories are divided into classes of \emph{passing particles} (with no reflections) and \emph{reflected particles} (with a~single reflection). The cases of \emph{trapped particles}, blocked between two mirrors at different sides of the source, are not considered: the first two passages these particles are taken into account in \emph{passing} and \emph{reflected} classes, while the following bounces are assumed to have a~negligible contribution to the total electron density in the energy range of our interest due to cooling.

    \item For each point $s$ along the MFL, we compute effective timescale and effective MF strength from eq.~\eqref{eq:effective_times_and_fields}. For \emph{passing} particles, the integration $0\to s$ is direct; for \emph{reflected} ones, the integral is divided into two parts: \mbox{$0\!\to\!s_m\!\to\!s$}, where $s_m$ is the corresponding mirror position.

    \item Effective timescale and MF strength allow for a~fast computation of the energy $E_0(\mu_\rm{obs}, s)$ of the particle at the moment of injection and the corresponding energy-evolution factor $\int_{0\to s}ds\,\partial_E\dot{E} /c\mu$. Assuming injection at the source to be static and isotropic, we compute the electron density $n(s) = dN/dE\,ds\,d\mu$ from
    \begin{equation}
        n(s) = \dfrac{Q(E_0)}{2|\mu_0 c|}
        \exp\left\{ -\intop_{0\to s} \dfrac{ds}{c\mu} \p_E\dot{E} \right\}.
        \label{eq:electron_density}
    \end{equation}

    \item Then we introduce a~regular grid along the projected coordinate $\zeta:~d\zeta = \sin\phi_\rm{obs}~ds$ and compute the projected electron density $n_P$:
    \begin{equation}
        n_P(\zeta) = \dfrac{dN}{dE~d\zeta~d\mu} = \dfrac{n\pbr{s(\zeta)}}{\sqrt{1 - \mu_\rm{obs}^2}}.
        \label{eq:projected_electron_density}
    \end{equation}

    \item From $n_P$, we recover the projected photon luminosity $l_P^\rm{phot}$ in gamma- and x-rays. To compute IC and synchrotron emission from the nebula we use analytical expressions~\cite{BlumenthalGould1970}.

    \item Finally, to construct 2D gamma-ray images, comparable to LHAASO significance maps, we integrate $l_P^\rm{phot}$ within the corresponding energy bins (\mbox{$25<\epsilon<100~\TeV$} and $\epsilon > 100~\TeV$) and add to the map a~cell-size-wide 2D Gaussian profile for each grid knot. The cell size is chosen to exceed the electrons' Larmor radii $<1~\pc$ below $1~\PeV$, but inferior to the LHAASO PSF ($\sim10~\pc$ at $3\dm{kpc}$). The resulting image is convolved with the energy-dependent PSF reported by the LHAASO collaboration.
    
\end{enumerate}

\noindent
This seven-step procedure leads to one- and two-dimensional count maps, which are compared to the LHAASO results.

\subsection{Ballistic simulation results}
\label{subsec:free-simulation-results}

We use the MFLs pregenerated for the GS turbulence study discussed in~\autoref{sec:GMF}. Each line is scaled according to the selected outer scale $L_\rm{max}\in\{200,~300,~400,~500\}\dm{pc}$, and rotated around the regular field direction to align the projected end of the Peanut with the $z$-axis; afterwards, the line is additionally rotated around the axis, connecting the pulsar and the endpoint, in order for the projected line to pass through the second hot spot.

\begin{figure*}[h!]
    \centering
    \includegraphics[width=\linewidth]{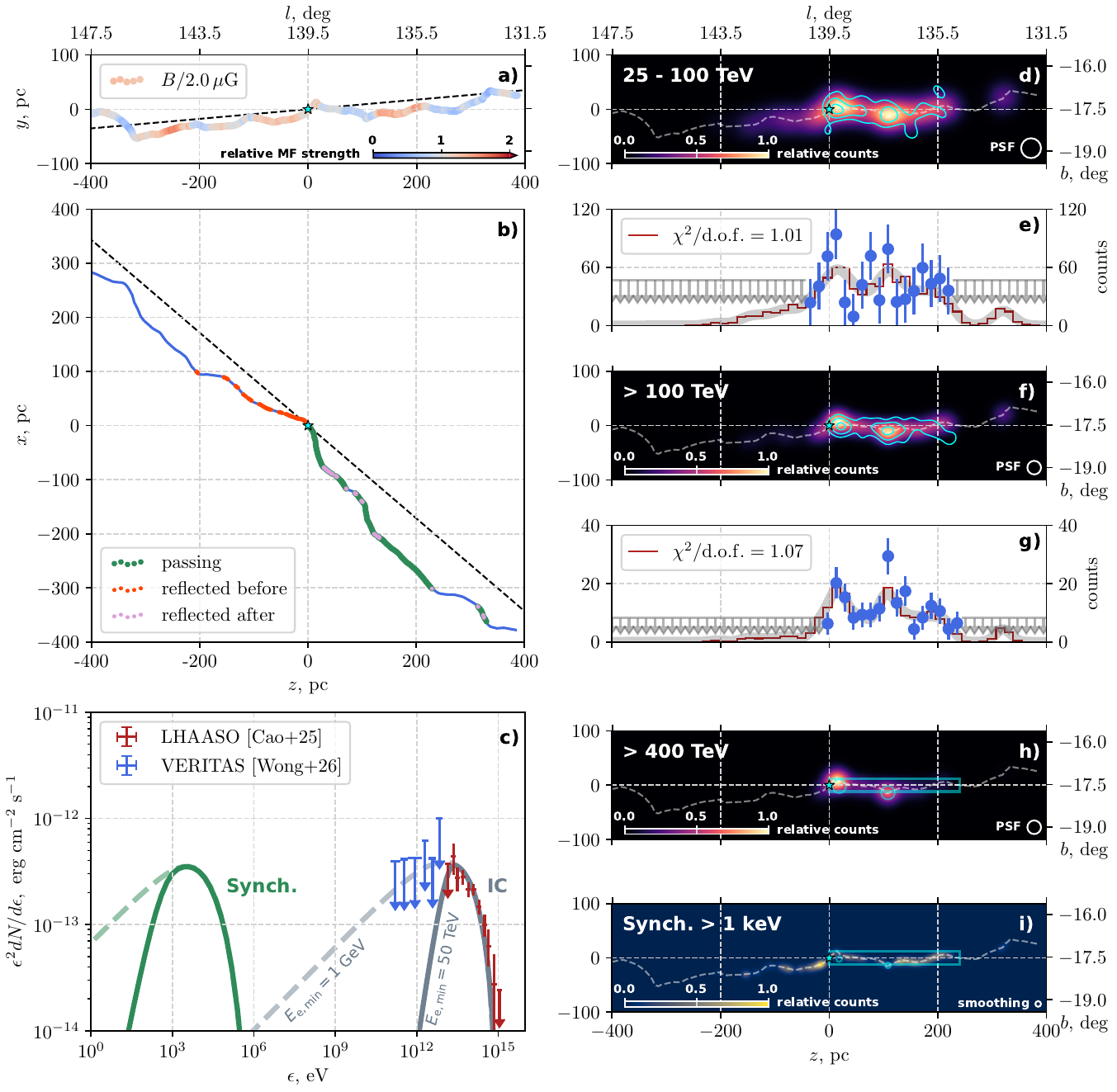}
    \vspace{-4ex}
    \caption{
        Simulation summary with parameters $B_\rm{tot} = 2~\uG$, $\eta = 0.7$, $L_\rm{max} = 300~\pc$, $\alpha_\rm{obs} = 50^\circ$ at the distance of 3~kpc.\\
        \textbf{(a):} MFL projection onto the sky ($zy$ plane, front-view) in absolute scales (pc) and in Galacic coordinates. The MF strength distribution normalized by the average MF strength of $2~\uG$ is showed in colors. Regular MF direction is shown with a black dashed line. The J0218 is marked with a~cyan star.\\
        \textbf{(b):} MFL projection onto the $zx$ plane (top-view), parallel to the Galactic plane. Green, red, and violet dots show particle classes responsible for the observed emission: \emph{passing}, \emph{reflected before}, and \emph{reflected after}, respectively. \\
        \textbf{(c):} Model predictions and measurements of the photon energy flux. Solid black and green lines correspond to IC and synchrotron fluxes with $E_{e,\min}=50~\TeV$, respectively. The dashed lines stand for a similar injection spectrum extending down to~\mbox{$E_{e,\min}=1~\GeV$}. Red points are the LHAASO measurements~\cite{Cao2026}; blue upper-limits are from VERITAS~\cite{Wong2026}. \\
        \textbf{(d):} Sky-projected gamma-ray counts map in the $25 < \epsilon < 100~\TeV$ energy bin. Cyan lines show LHAASO significance contours for 3, 5, and $7\sigma$~\cite{Cao2026}. White circle shows the adopted PSF size. The 2D image is normalized by the maximal flux. \\
        \textbf{(e):} 1D profile along the $z$-axis in the $25 < \epsilon < 100~\TeV$ energy bin. Data points (blue) are 1D LHAASO profile measurements, upper-limits (gray) are derived from point errors as discussed in the text. A red line shows the model predicted photon flux (PSF is taken into account).\\
        \textbf{(f)-(g):} Same as \textbf{(d)}, \textbf{(e)}, but in the energy bin of $\epsilon > 100~\TeV$. Cyan LHAASO contours are shown for 3, 6, and $9\sigma$.\\
        \textbf{(h):} Same as \textbf{(d)}, but the energy is above $400~\TeV$. Cyan rectangle corresponds to the LHAASO strip; the circles point hot spot positions. PSF size is the same as in panel~(f).\\
        \textbf{(i):} Same as \textbf{(d)}, but for synchrotron photons with energies above $1~\keV$. The smoothing of $0.1^\circ$ is shown instead of the PSF.
    }
    \label{fig:best_ballistic_m}
\end{figure*}

For each scaled line, we propagate particles and compute corresponding IC and synchrotron emission using the algorithm established in~\autoref{subsec:free-particle-propagation} for a~selection of observation angles $\alpha_\rm{obs} \in \{ 30^\circ\!\!,~40^\circ\!\!,~...,~90^\circ \}$ and magnetic fields $B_\rm{tot}\in\{1.0,~2.0,~4.0\}~\uG$. Injection spectrum $Q(E)$ is selected to be the LHAASO best-fit electron spectrum~\cite{Cao2026}. To find the optimal MFL configuration, we calculate 1D profiles along the $z$-axis, corresponding to the Galactic parallel $b=-17.5^\circ$ and compared it to the data. As LHAASO 1D profiles are binned in right ascension (RA), we change bin width by taking into account an angle of $13.5^\circ$ between the Galactic and Equatorial parallels at the source.

Assuming errors in LHAASO's 1D profiles are background dominated, for $25 < \epsilon < 100~\TeV$ and \mbox{$\epsilon > 100~\TeV$}, we estimate expected background rate in each 1D bin and thus compute $2\sigma$ upper limits for counts outside of the LHAASO measurement region. Data points with values below $1\sigma$ are also substituted with these $2\sigma$ upper-limits. For the model in each of the two energy-bins, we introduce independent normalization parameters, which maximize the corresponding likelihood. In total, we studied around 1\,000 different MFL configurations for each set of parameters ($\eta$, $L_\rm{max}$, $\alpha_\rm{obs}$, $B_\rm{tot}$), which resulted into less than 20 well-fitted configurations ($\chi^2/\rm{d.o.f.} < 1.4$ in the $\epsilon > 100~\TeV$ bin) per set. The most frequent configurations correspond to $L_\rm{max} = 300~\pc$ and $400~\pc$, $30^\circ < \alpha_\rm{obs} < 60^\circ$ and $B_\rm{tot} = 2~\uG$. The results of the parameter space study are summarized in~\autoref{fig:ballistic_limits}.

\autoref{fig:best_ballistic_m} shows one of the best-fit configurations with $B_\rm{tot} = 2~\uG$, $\eta = 0.7$, $L_\rm{max} = 300~\pc$, $\alpha_\rm{obs} = 50^\circ$. The 1D profiles are fit well with the LHAASO data points in both energy bins with corresponding $\chi^2/\rm{d.o.f.}$ values of $1.01$ and $1.07$, respectively. In the maps, location and brightness of both hot spots is well reproduced. Further away along the $z$-axis, a~weaker spot is seen; its total photon counts lies well below the upper-limits. Due to synchrotron cooling, this spot is not present at the high-energy photon map ($\epsilon > 400~\TeV$). 

The physical origin of the first and second hot spots lies in geometry of the MFL: at $z\sim20~\pc$, the turbulent mode decreases the local angle of the MFL with respect to the LOS. This amplifies the emission due to two factors:
\begin{enumerate}
    \item The number of radiating particles per unit axis length increases, as seen from~\eqref{eq:projected_electron_density};
    \item The smaller observed pitch angle means reduced synchrotron losses during particle propagation from the pulsar to the observed location.
\end{enumerate}

The length of the Peanut is also constrained by geometric reasons: around the endpoint, $z=240~\pc$, the turbulent modes curve the MFL away from the observer and thus make the electron population ``invisible''.

We note a~remarkable difference between the IC and synchrotron maps shown in panels~(d), (f), (h), and (i) of~\autoref{fig:best_ballistic_m}. While IC scattering on CMB is pitch-angle-independent, synchrotron intensity and its characteristic frequencies are proportional to $\sin\phi_\rm{obs}$. This makes regions of enhanced gamma-ray signal (which have smaller $\phi_\rm{obs}$) faint in synchrotron, and, contrarily, weaker gamma-ray regions appear brighter in~x-rays.

In panel~(c) of the~\autoref{fig:best_ballistic_m}, we show the LHAASO and VERITAS measurements compared to predicted spectra of the gamma-ray and x-ray emission. For a demonstration, we select two different low-energy (LE) cutoffs: $E_{e,\rm{min}} = 50~\TeV$, which is the highest LE cutoff still fitting the gamma-ray data, and $E_{e,\rm{min}} = 1~\GeV$, which also fits the LHAASO data well and agrees with the VERITAS upper-limits. The LE cutoff of the soft PLSuperEC~\cite{Cao2024} electron injection spectrum ($p=-2.48$, $E_\rm{cut}=770~\TeV$, $\beta = 2$) determines the total power pumped into the electron population. For LE cutoffs below $0.5~\TeV$, the total power required to sustain the gamma-ray radiation from a free-propagating electron population exceeds 80\% of the pulsar spin-down luminosity~$\dot{\mathcal{E}}$. This places additional constraints on the injected spectrum.

Similarly to~\autoref{sec:diffusion}, we leave the question of detectability of the x-ray signal for further investigation.

\subsection{Influence of reflected particles}

Another noticeable prediction of our modeling is the presence of gamma-ray tails on the west of the pulsar, which are not observed at a significant level by LHAASO. These tails are produced by the \emph{reflected} particles (reflected once by a~magnetic mirror). Panel~a) of~\autoref{fig:best_ballistic_m} shows two regions of enhanced MF strength $B$: at $-300 < z < -200~\pc$ and at $100 < z < 200~\pc$. These regions act as magnetic mirrors for electrons with larger pitch-angles, making them reflect and contribute to the visible gamma-ray signal. Reflected particles have higher pitch-angles and spend more time before reaching the observed point. Thus they cool more strongly and have smaller contribution to gamma-ray signal at higher energies. If at $25 < \epsilon <100~\TeV$ the tail in~\autoref{fig:best_ballistic_m} is relatively visible, it becomes less significant at $\epsilon > 100~\TeV$ and disappears completely above~$400~\TeV$.

For a~particle injected with a~pitch-angle $\mu_0$, the mirror positions are determined by the MF strength distribution along the MFL and, most importantly, MF strength $B_0$ at the source. As our selection of MFL center positions was uniformly distributed in the pregenerated MF cube, the MF strength at the center (at the source) is usually lower than the its average strength along the line. This is supported by the fact that MFLs tend to go to stronger field regions due to magnetic flux conservation.

\begin{figure}[t!]
    \centering
    \includegraphics[width=\linewidth]{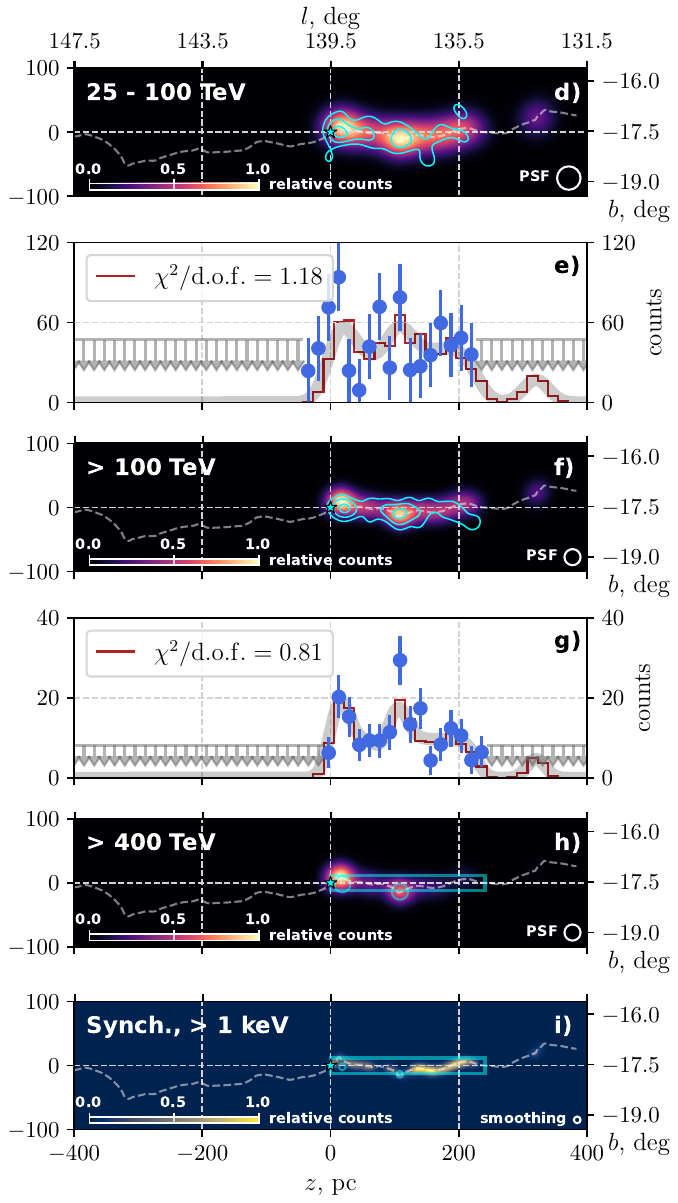}
    \vspace{-4ex}
    \caption{
        Gamma-ray and synchrotron maps with relevant 1D~profiles for \emph{passing} electron population only. Panels are similar to panels \textbf{(d)}...\textbf{(i)} of~\autoref{fig:best_ballistic_m}.
    }
    \label{fig:best_ballistic_noref}
\end{figure}

At the same time, the actual field strength at the injection position, which we consider located somewhere in the PWN, may differ significantly from the GMF strength. Due to $\mathcal{I}_0$ conservation, an increased $B$ would lead to a~decrease in number of reflected particles, however also collimating the injected particle distribution. To demonstrate the influence of reflected particles, we show~\autoref{fig:best_ballistic_noref}, in which only \emph{passing} particles contribute to the gamma-ray emission. While conserving the main features, the source becomes strictly one-sided. Relative brightness of the synchrotron map shifts from the vicinity of the source to the western edge of the Peanut, where magnetic field strength in this particular MFL realization is elevated.

Further gamma-ray and x-ray observations, combined with more detailed studies of the PWN powered by J0218, may shed light on particle transport in the injection region. Under the assumption of ballistic propagation, this would help determine the transport picture between the two limiting cases considered here: \emph{passing} particles alone and \emph{passing} particles supplemented by \emph{reflected} ones.

    \section{Discussion}
\label{sec:discussion}

From a handful of sources LHAASO observed outside the Galactic plane, only two are significantly extended: V4641~Sgr and Peanut. Such extended gamma-ray sources are prospective laboratories for studying the GMF at their locations. In particular, a source extending over several hundred parsecs can indicate the direction of the local large-scale~GMF, while its perpendicular extension may constrain turbulence properties~\cite{Kleimenov2026}. 

In this work, we model the gamma-ray emission from the Peanut source~\cite{Cao2026}, which was reported in two energy bands: 
$25<\epsilon<100$ TeV, $\epsilon>100$ TeV, and in separate photons above $400$ TeV. Following the association proposed by LHAASO~\cite{Cao2026}, we assume that the millisecond pulsar PSR~J0218+4238 powers the gamma-ray emission from the Peanut. The chosen distance of~$3\dm{kpc}$ is based on several independent measurements; at this distance, the Peanut length is around $240~\pc$.

\entry{Kolmogorov (diffusion) scenario}

In Kolmogorov turbulence, electrons diffuse along the principal magnetic field line. In this regime, from cooling constraints, we found that the total GMF cannot exceed $7~\uG$, and the diffusion coefficient normalization has to lie within an order of magnitude from the typical value of $D_\rm{\prl,1\PeV} = 10^{30}\dm{cm^2~s^{-1}}$. We note that in the diffusion case, stronger cooling constraints are associated with the observation of $100~\TeV$ photons at the eastern edge of the Peanut, rather than with the $760~\TeV$ photon in the middle of the strip. The fraction of the pulsar's spin-down luminosity required to power the electron distribution above $50~\TeV$ lies between 0.4\% and 10\%, increasing with the MF strength and peaking at the observed angle of $90^\circ$. Typical values of the required power remain around~1\%. 

Our simulations of electron propagation in a turbulent MF are based on the solution of the motion equation for each particle in the stacked MF grids. This approach allows for a demonstration of the possibility to describe the Peanut morphology in the diffusion scenario. The best-fit configuration we obtain has the following parameters:  $L_\rm{max} = 500~\pc$, $\eta=1.0$, $B_\rm{tot} = 1~\uG$, $\alpha_\rm{obs} = 112^\circ$. Our simulations indicate that the primary mechanism for the Peanut's hot spots formation in Kolmogorov turbulence may be magnetic bottling: in~\autoref{fig:diffusive_bestfit}, we see a link between the gamma-ray emission peaks and the MF strength along the primary MFL minima. Such an effect of magnetic trapping of particles may indeed exist in realistic MHD turbulence as suggested in Ref.~\cite{XuLazarian2020}. However, the structure of the field line suggests that MFL geometry features may also cause the appearance of hot spots.

While generally attractive, the Kolmogorov scenario has an embedded difficulty. To reproduce the strongly asymmetric morphology of the Peanut, we need to assume one-sided particle injection and place a magnetic mirror at the source location, to prevent particles from diffusing back to the other side of the source and produce gamma-ray emission on both sides. Such assumptions can however naturally be satisfied. First, asymmetric particle escape from PWNe has already been suggested by several theoretical studies, for example due to the geometry of the PWN magnetic fields~\cite{2018MNRAS.480.5419B}, or to magnetic reconnection between the PWN and interstellar magnetic fields~\cite{2019MNRAS.485.2041B}. Second, it is reasonable to expect a magnetic mirror to form at the source location: the PWN pushes on the surrounding interstellar medium, and interstellar MFLs are expected to be draped around the tail of the nebula. This should therefore create regions of stronger interstellar magnetic fields around the nebula, and may create a magnetic mirror at the source location. In that sense, the Kolmogorov scenario is not contrived despite the need for these two additional assumptions to reproduce the Peanut morphology.

In comparison to the GS turbulence, the use of computationally heavy particle propagation simulations does not allow for exploration of significant parts of the parameter space and our best-fit configurations exist for a particular set of global parameters. This can be improved by developing a simplified treatment of particle propagation in Kolmogorov turbulence. The complication lies in the necessity to include magnetic bottling, which is influenced by pitch-angle scattering. A more detailed study of this case would be needed.

\entry{Goldreich-Sridhar (ballistic) case}

In the case of GS turbulence, electrons propagate in the ballistic regime. Cooling constraints are dominated by the highest-energy $760~\TeV$ photon located at the projected distance of $100~\pc$ away from the pulsar. They constrain magnetic field to be less than $8~\uG$ except several improbable cases with $\alpha_\rm{obs} < 30^\circ$. The power to feed the electron population above $50~\TeV$ lies between $4\%$ and $8\%$ for the range of angles between $15^\circ$ and $75^\circ$. This, as expected, exceeds the diffusion regime power requirements.

Ballistic particle propagation is modeled in the ultrarelativistic limit of the guiding-center approximation with radiative losses. In this setting, electron motion and energy evolution are decoupled, which allows for a simplified treatment of the problem and for an investigation of a broader parameter space region. The best-fit configuration has the following parameters: $L_\rm{max} = 300~\pc$, $\eta=1.0$, $B_\rm{tot}=2~\uG$, $\alpha_\rm{obs} = 50^\circ$. \autoref{fig:best_ballistic_m} confirms that in the ballistic case the hot spots originate from the MFL geometry. Both the first and second Peanut hot spots correspond to the regions of the line oriented towards the observer, which both increases the electron projected density and selects the particle with smaller pitch-angle, which were less affected by cooling during their propagation. The study of the parameter space we conducted shows that most of the best-fit configurations have observed angles between $30^\circ$ and $60^\circ$ (see~\autoref{fig:ballistic_limits}). The $2~\uG$ magnetic field is the most frequent among the best configurations.

In contrast to the Kolmogorov scenario, our treatment of the GS case does not allow for the formation of local magnetic bottles with enhanced radiation intensity. Due to conservation of $\mathcal{I}_0$ for each particle (see eq.~\eqref{eq:final-evolution}), magnetic trapping is allowed only on large-scales on two different sides from the source. This, however, does not exclude the possibility to form magnetic-bottle-induced hot spots in a more realistic picture of GS turbulence. To study this phenomenon, a detailed simulation of particle propagation similar to our treatment of Kolmogorov turbulence in the GS scenario is needed.

As discussed above for the Kolmogorov scenario, the interstellar magnetic field strength in the immediate vicinity of the source may be locally stronger due to the PWN pushing on the surrounding medium. Such an effect is not included in our simulations for ballistic particle propagation. Hence, in practice, particles escaping from the PWN may be injected in a region with stronger magnetic fields than in our simulations. This would have the effect of reducing the amount of reflected particles in our calculations in the previous Section, and would therefore make our predicted gamma-ray emission even more asymmetric, further suppressing the weak tail on the western side of the pulsar in \autoref{fig:best_ballistic_m}.

\begin{figure}[b!]
    \centering
    \includegraphics[width=\linewidth]{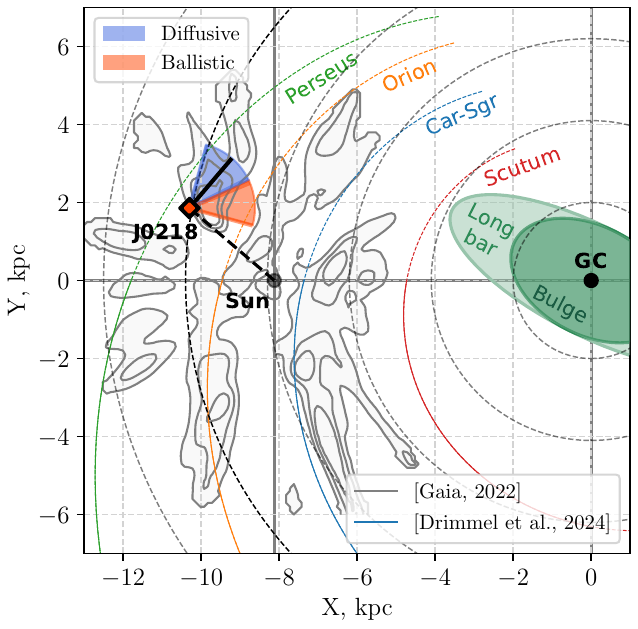}
    \caption{
        Relative position of the J0218 in the Galactocentric coordinates. Black line shows the projected direction of the Peanut. Preferred diffusion observation angles $65^\circ < \alpha_\rm{obs} < 115^\circ$ are shown with a blue wedge. A red wedge demonstrates preferred observation angles for the ballistic scenario $30^\circ < \alpha_\rm{obs} < 65^\circ$. Representation of spiral arms relies on results from Gaia DR3 (black contour plot)~\cite{Gaia_2022} and Cepheid-based parametrization (colored lines)~\cite{Drimmel2024}. Galactocentric coordinate parameters are taken from \texttt{Astropy v4.0}~\cite{Astropy2022}; the spiral arms are plotted with the use of the~\texttt{SpiralMap} Python library~\cite{Prusty2026}.
    }
    \label{fig:location}
\end{figure}

\entry{Implications for the Galactic magnetic field}

Kolmogorov and GS turbulence scenarios give rise to curious but distinct conclusions about the direction of the regular field at the location of the Peanut. While analytical constraints on observer angles in both cases are relatively weak, the angles preferred by simulations are different, as shown in~\autoref{fig:location}.

For the Kolmogorov scenario, several best-fit angles lie within $65^\circ < \alpha_\rm{obs} < 115^\circ$, which is close to the direction orthogonal to the observer. We recall that in the diffusion scenario, the two emission cases with the regular field oriented at $\alpha_\rm{obs}$ and \mbox{$180^\circ - \alpha_\rm{obs}$} are identical. If confirmed, the Kolmogorov case opens two possible scenarios, depending on the position of the Peanut source in the magnetic arm in the halo. At $Z=-0.9\dm{kpc}$, the Peanut lies on the border between those structures. In the former case, the strip direction should be close to orthogonal to the LOS, parallel to Perseus arm in KST24 model~\cite{Korochkin:2024yit} (solid black line in~\autoref{fig:location}). In the latter scenario, the filament should extend along a large circle centered at the Galactic center (thin dashed black line in~\autoref{fig:location}).

For GS turbulence, our parameter-space study shows a preference for acute angles from $30^\circ$ to $60^\circ$. If the GS scenario is confirmed, it will challenge present day GMF models, which do not predict lines with a small direction angle to the Earth in this part of the Galaxy.

\entry{Distinguishing between the scenarios}

With the current gamma-ray data, Kolmogorov and GS scenarios are indistinguishable. However, due to the high gamma-ray flux from the Peanut source, experimental data quality will be increasing with time, which will bring better understanding of the properties of the source.

A feature specific to the Kolmogorov scenario is the width of the particle distribution induced by perpendicular diffusion. Resolving this width is possible by either gamma-ray or x-ray facilities. Long exposure x-ray measurements of the filament may reveal synchrotron emission hot spots slightly shifted forward from the gamma-ray enhancements. In addition, magnetic-bottle-like confinement of electrons might be energy-dependent. Existing indications in the LHAASO data, corresponding to the position of the first hot spot, are not decisive and lie within the PSF size. Potential revelation of energy-dependent morphology would support magnetic-bottling over geometric hot spot formation mechanism.

In the Goldreich-Sridhar scenario, the Peanut scales are constrained by the magnetic field line geometry, which implies a probable appearance of another spot after the currently visible end of the Peanut. The actual length of the electron population is defined by the total magnetic field and in the case of $1~\uG$ is predicted to be much longer than the observed filament. We may also observe weak emission below $100~\TeV$ on the western side of the source due to particles reflected due to global magnetic mirroring. Contrary to Kolmogorov scenario, our x-ray predictions do not follow the gamma-ray hot spots. Due to predicted geometric origin of the hot spots in GS turbulence, x-ray signal is enhanced in the cold spots of the gamma-ray signal. This alternation is, however, typical to geometrically produced hot spots, and may appear in both turbulence scenarios. Finally, the emission direction, being significantly different from existing large-scale GMF predictions, might become decisive with the development of GMF models.

\entry{Further development}

The questions of x-ray emission detectability, exploration of the broader parameter space in the Kolmogorov scenario, and detailed simulations of the GS turbulence are potential directions of further theoretical progression.

In terms of experimental studies, the LHAASO collaboration data are being accumulated, and in a few years the statistics will double, leading to a better understanding of morphological properties. Located in the Northern sky, the Peanut is a valuable target for the next generation of IACTs: excellent angular resolution of CTA North~\cite{CTAO_Book2019, Wong2026} and LACT~\cite{Zhang2026} in complementary energy ranges will provide important information about the morphology of the source.

Despite the large angular size of the source, the Peanut remains an interesting target for x-ray facilities. The x-ray emission, expected to extend up to 100~keV (see~\autoref{fig:best_ballistic_m}) has a predicted total flux between $10^{-13}$ and $10^{-12}\dm{erg~cm^{-2}~s^{-1}}$ in a $2~\uG$ field.  The x-ray detectability is determined by the real width of the filament, which is currently bounded by the electron Larmor radius from below and by the LHAASO PSF from above. Upper-limits for the x-ray flux, which could be obtained with INTEGRAL data~\cite{Winkler2003} or observations with CXO~\cite{Weisskopf2000} and XMM-Newton~\cite{Jansen2001} will provide useful constraints on particle propagation and GMF properties. Very recent x-ray observations of the gamma-ray hot spot regions by the Einstein Probe telescope~\cite{Yuan2025} resulted into flux upper-limits of $\sim5\times10^{-14}\dm{erg~cm^{-2}~s^{-1}}$ for each hot spot~\cite{Feng2026}, which is in line with our findings.

    \section{Conclusion}
\label{sec:conclusion}

In the present work, we provide examples showing that the LHAASO's Peanut source may be reproduced via inverse Compton scattering of CMB photons by high-energy electrons, propagating along the lines of the Galactic magnetic field. The high-energy electrons may be injected by the millisecond pulsar PSR~J0218+4232, identified by LHAASO as the sole candidate to supply sufficient power for the Peanut region. We find that a substantial part of its spin-down luminosity (from 0.5\% to 10\%) is required to power the electron population above~$50~\TeV$. Together with the maximal possible energy around $1~\PeV$, this implies a very efficient acceleration mechanism at the Pulsar.

The Galactic magnetic field, which consists of the regular and turbulent components, is modeled here in the two limiting scenarios: Kolmogorov turbulence, corresponding to diffusive particle propagation, and Goldreich-Sridhar turbulence, in which electrons propagate ballistically. The remarkable straightness of the Peanut region places constraints on the turbulent field: we find that corrected by the unknown direction of the regular field at the location of the Peanut, the probability to reconstruct a straight filament with projected length of 240~pc is below $70\%$ and $30\%$ for turbulent/regular ratios $\eta = 0.5$ and $\eta=1.0$, respectively (\autoref{fig:line_acceptance_all}). This fraction increases with the observed angle, as the physical scale of the Peanut filament decreases.

Kolmogorov and the Goldreich-Sridhar turbulence correspond to diffusive and ballistic propagation, respectively. We derived analytical constraints for both models of turbulence, relying on cooling, transport and power constraints. In both scenarios, the magnetic field cannot exceed $\sim8~\uG$. The observation angles in the diffusive regime are constrained to be between around $30^\circ$ and $150^\circ$, while simulations prefer values closer to $90^\circ$; our best-fit scenario corresponds to $112^\circ$ (see~\autoref{fig:diffusive_bestfit}).
Ballistic propagation, on the contrary, can in principle represent the Peanut only for acute observed angles. Parameter space study of this case prefers angles from $30^\circ$ to $60^\circ$; our best-fit scenario corresponds to the observed angle of $50^\circ$ (see~\autoref{fig:best_ballistic_m}).

Our simulations reveal two potential mechanisms of hot spot formation: in the Kolmogorov scenario, we observe magnetic-bottle-like confinement of particles in the region of lower magnetic field strength; in GS turbulence, geometric projection effect of large-scale MFL fluctuations is responsible for accumulating more electrons along the LOS. While we provide examples where these mechanisms give rise to hot spots in Kolmogorov and GS turbulence, we cannot conclude that there is a one-to-one correspondence between the hot spot formation mechanism and the type of turbulence. A deeper parameter space study for Kolmogorov turbulence and more detailed modeling of the GS case are necessary to explore this question.

The LHAASO's Peanut source will play an important role in cross-checking the GMF models. In the case of GS turbulence, as seen from~\autoref{fig:location}, the preferred direction of the regular magnetic field has an acute angle with respect to the LOS, which is not predicted by any of the existing models. For the Kolmogorov scenario, the regular MF direction may be consistent either with the main field direction in the Perseus arm in the KST24 model, or with the toroidal MF of the Galactic halo present in most of modern descriptions of the GMF.

The Peanut is a promising candidate for further gamma-ray and x-ray studies, and will serve in the future as an excellent laboratory for particle propagation in the Galaxy.

    \section*{Acknowledgements}

Work of D.S. has received partial support under the program
`` Investissement d’Avenir '' launched by the French Govern-
ment and implemented by ANR, with the reference `` ANR-
18-IdEx-0001 '' as a part of its program  ``~Emergence~''. 

    \bibliography{lit}
    
    \appendix
    \section{Properties of J0218+4238}
\label{appendix:pulsar}

In this section, we summarize the properties of PSRJ0218+4238 and of the Peanut gamma-ray region assuming its association to the pulsar.

\entry{The controversy in J0218 distance measurements}

The pulsar was detected in~1995, its distance was estimated above $5.7\dm{kpc}$~\cite{Navarro1995} based on existing Galactic density models (TC93)~\cite{Taylor1993}. Assuming this distance, J0218 became the most luminous pulsar ever observed.

In~2003, the 10-m optical Keck telescope was able to resolve the WD component of the system~\cite{Bassa2003}. Based on WD cooling age, which the authors of~\cite{Bassa2003} assumed to be similar to the Pulsar age, they placed a constraint of $2.5 < d < 4~\dm{kpc}$, which contradicted initial DM-based estimates. In parallel, a new electron density model (NE2001) was issued, which predicted the distance of $2.7\dm{kpc}$~\cite{Cordes2003}.

The first direct distance measurement through parallax was made by~\cite{Du2014} with European VLBI Network in 2014 and resulted in $6.3^{+8.0}_{-2.3}\dm{kpc}$, which was consistent with first TC93-based estimates. As mentioned previously, this distance made the pulsar the most luminous ever observed and challenged standard pulsar models. Shortly, the tension was relieved by a work which pointed out that distance measurements were dominated by the Lutz-Kelker bias~\cite{Verbiest2014}. The revised distance turned out to be $3.15^{+0.85}_{-0.60}\dm{kpc}$, which was also consistent with the DM~(NE2001) and WD distances. A more recent electron density model (YMW16) confirmed the result by giving the distance of $2.9\dm{kpc}$~\cite{Yao2017}. \emph{Based on these results, we assumed the distance to be~$3\dm{kpc}$. }

Nevertheless, we have to note that in 2024, CHIME reported an indirect distance estimation based on Doppler shift variations attributed to Shklovskii acceleration~\cite{Tan2024}. Their estimate gives $d = 6.7\pm0.7\dm{kpc}$. All the aforementionned distance estimations and measurements are summarized in~\autoref{tab:distances}.

\begin{table}[h!]
    \vspace{-4ex}
    \centering
    \caption{
        Distance measurements to J0218 via different methods (WD -- white dwarf, DM -- dispersion measure).
    }
    \begin{tabular}{c|c|c|c}
        Method  & $d,\dm{kpc}$ & Ref. & Date \\
        \hline
        VLBI             & $6.3^{+8.0}_{-2.3}$    & \cite{Du2014}       & Feb~2014\\
        VLBI (corrected) & $3.15^{+0.85}_{-0.60}$ & \cite{Verbiest2014} & Jul~2014 \\
        \hline
        WD companion cooling & $2.5 < d < 4$ & \cite{Bassa2003} & 2003 \\
        \hline
        DM~(TC93~\cite{Taylor1993}) & $>5.7$ &\cite{Navarro1995} &  1995\\
        DM~(NE2001~\cite{Cordes2003}) & $2.7$ & \cite{Bassa2003} & 2003 \\
        DM~(YMW16~\cite{Yao2017}) & $2.9$ & \cite{Yao2017}  & 2017 \\
        \hline
        Shklovskii acceleration & $6.7\pm0.7$ & \cite{Tan2024} & 2024 
    \end{tabular}
    \label{tab:distances}
\end{table}

\entry{The companion star's influence is negligible}

The binary orbit has a semi-major axis of approximately \mbox{$2$~lt-s} ($6\times10^5\dm{km}\approx1.9\times10^{-8}\dm{pc}$)~\cite{Navarro1995, Acciari2021}, which is nearly seven orders of magnitude smaller than the expected stand~off distance. Consequently, the companion star resides well inside the free pulsar wind region and has a negligible effect on the termination shock structure.

\begin{table}[h!]
    \centering
    \caption{
        Summary of J0128: 
    }
    \label{tab:summary}
    \begin{tabular}{c|c|c}
        Quantity    & Value & Ref. \\
        \hline
        (RA, Dec), deg     & $34.526$, $42.538$ & \multirow{2}{1em}{\cite{Navarro1995}} \\
        $(l,~b)$, deg\footnote{Derived from \cite{Navarro1995} using \texttt{astropy}~\cite{Astropy2022}: \href{https://www.astropy.org}{astropy.org}}   & 139.508, -17.527 & \\
        \hline
        $\mu_\alpha\cos\delta,\dm{mas~yr^{-1}}$ & $5.35\pm0.05$  & \multirow{2}{1em}{\cite{Du2014}} \\
        $\mu_\delta,\dm{mas~yr^{-1}}$           & $-3.74\pm0.12$ & \\
        \hline
        $\gamma_r,\dm{km~s^{-1}}$ & $-57\pm52$ & \cite{Bassa2003} \\
        \hline
        $P,\dm{ms}$ & 2.32 & \cite{Navarro1995} \\
        $\dot{E},\dm{erg~s^{-1}}$ & $2.4\times10^{35}$ & \cite{Acciari2021}\footnote{In the 3PC~\cite{Smith2023a} $\dot{E} = 4\pi^2I_0\dot{P}P^{-3}$ is calculated in the assumption of the neutron star momentum of inertia of $I_0 = 10^{45}\dm{g~cm^2}$, which corresponds to a homogeneous sphere of mass $1.3M_\odot$ and radius $10~\rm{km}.$} \\
        \hline
        $F_\rm{LAT},\dm{erg~cm^{-2}~s^{-1}}$ & $(48.0\pm1.2)\times10^{-12}$ &
        \multirow{2}{1.5em}{\cite{Smith2023a}} \\
        $L_\rm{LAT},\dm{erg~s^{-1}}$ at $3\dm{kpc}$ & $5.2\times10^{34}$ &
    \end{tabular}
\end{table}

\begin{table}[h!]
    \vspace{-2ex}
    \centering
    \caption{
        The Peanut summary~\cite{Cao2026} with angular and volumetric values ($d=3\dm{kpc}$) for
        the strip size $\Delta$, distances ~$\ell_l$ and ~$\ell_b$ from J0218 along $l$ and $b$, respectively, the energy flux $F_{13}$ in $10^{-13}\dm{erg~cm^{-2}~s^{-1}}$, and luminosity $L_{32}$ in $10^{32}\dm{erg~s^{-1}}$.
    }
    \label{tab:peanut_with_j0218}
    \begin{tabular}{c|c|c}
           & Angular & Volumetric ($d=3\dm{kpc})$ \\
         \hline
         \multirow{3}{*}{Strip} & $\Delta_l = 4.60^\circ$ & $\Delta_l = 240~\pc$ \\
                                & $\Delta_b = 0.46^\circ$ & $\Delta_b = 24~\pc$  \\
                                & $F_{13} = 5.3 $ & $L_{32} = 5.7$ \\
         \hline
         \multirow{3}{*}{First hot spot}& $l=139.14^\circ$ & $\ell_l = -18.4~\pc$ \\
                                        & $b=-17.57^\circ$ & $\ell_b = -2.1~\pc$ \\
                                        & $F_{13} = 1.06$ & $L_{32} = 1.14$ \\
         \hline
         \multirow{3}{*}{Second hot spot}   & $l=137.34^\circ$ & $\ell_l = -108.4~\pc$ \\
                                            & $b=-17.79^\circ$ & $\ell_b = -13.7~\pc$  \\
                                            & $F_{13} = 0.91$ & $L_{32} = 0.98$ \\
         \hline
         PSF, 25 - 100 TeV        & $0.35^\circ$  & $18~\pc$ \\
         PSF, $> 100~\TeV$        & $0.24^\circ$  & $13~\pc$ \\
    \end{tabular}
\end{table}

Relevant properties of J0218 are summarized in~\autoref{tab:summary}; Peanut morphological features are calculated at the distance of~\mbox{$d=3\dm{kpc}$} and presented in~\autoref{tab:peanut_with_j0218}.

    \section{Magnetic field generation}
\label{appendix:mf-generation}

To generate spatial 3D turbulent cubes of the magnetic field, we separate regular $\v{B}_0 = (0,0,B_0)$ and turbulent $\delta \v{B}$ components. The turbulent field is generated in Fourier space on a 3D grid of size $256^3$ with Gaussian point statistics, each mode is weighted with the turbulence power spectrum
\begin{equation}
    \delta\tilde{b}_j = \mathcal{N_C}(0, 1) \sqrt{\mathcal{P}(\v{k})} + \mathrm{h.c.},
\end{equation}
where $\delta\tilde{b}_j$ is the $j$-th component of the non-dimensional turbulent MF in the $k$-space; $j \in \{k_x,k_y,k_z\}$, $\mathcal{N_C}$ stands for complex normal distribution.

For Kolmogorov turbulence~\cite{Kolmogorov1941},
\begin{equation}
    \mathcal{P}(\v{k}) \propto k^{-11/3}~\boldsymbol{1}(k_\rm{min} < k < k_\rm{max}),
\end{equation}
where $k = |\v{k}|$, and $\boldsymbol{1}(A) = \{1 \text{ if A is True, else } 0\}$.

The lower boundary $k_\rm{min}$ determines the \emph{outer scale} of the turbulence $L_\rm{max} = 2\pi k_\rm{min}^{-1}$.

For GS turbulence, we tested both common realizations: with a window function~\cite{Chandran2000} and an exponential cutoff~\cite{Yan2002}. As the differences in realizations are expected to affect mostly small scales, we saw no substantial difference between the two and decided to stick to the latter realization, which has the power spectrum
\begin{multline}
    \mathcal{P}(\v{k}) \propto k_\perp^{-10/3}\exp\left\{-(2\pi)^{1/3}k_\prl k_\perp^{-2/3} k_\rm{min}^{-1/3}\right\} \times \\
    \times \boldsymbol{1}(k_\rm{min} < k < k_\rm{max}),
\end{multline}
where $k_\prl = k_z$ and $k_\perp = \sqrt{k_x^2 + k_y^2}$.

The spatial turbulent component $\delta \v{b}$ is an inverse Fourier transform of $\delta\tilde{\v{b}}$, normalized in such a way that
$\sqrt{\lrangle{(\delta \v{b})^2}} = 1$, where $\lrangle{\cdot}$ stand for spatial average.

To add regular component $\v{b}_0 = (0, 0, 1)$, we select the turbulent/regular ratio $\eta$ and define the total MF
\begin{equation}
    \v{B} = B_\rm{ampl}~\dfrac{\v{b}_0 + \eta \,\delta\v{b}}{\sqrt{1 + \eta^2}},
\end{equation}
such that $\sqrt{\lrangle{\v{B}^2}} = B_\rm{ampl}$, where $B_\rm{ampl}$ is the desired magnetic field strength.

For Kolmogorov turbulence, we use the notion of \emph{coherence length}. Following Ref.~\cite{2002UHE}, the coherence length $\mathcal{L}_c$ of an isotropic turbulent field can be written as
\begin{subequations}
    \begin{align}
        \label{CorrelationLength}
        \mathcal{L}_c &=
        \frac{L_{\max}}{2}
        \frac{\alpha - 1}{\alpha}
        \frac{1-(L_{\min}/L_{\max})^{\alpha}}
        {1-(L_{\min}/L_{\max})^{\alpha-1}},\\
        \mathcal{L}_{c,\rm{Kolm}} &\approx L_\rm{max} / 5.
    \end{align}
\end{subequations}

In the particle propagation simulations of~\autoref{sec:diffusion}, we use stacked grids to describe the magnetic field on a wide range of scales. When the turbulence is distributed among nested grid levels, the magnetic energy assigned to the \(i\)-th level is proportional to 
\begin{equation}
    B_{{\rm {rms}},i}^2
    \propto
    B_{\rm {rms}}^2
    \frac{L_{i+1}^{\alpha-1}-L_i^{\alpha-1}}
    {L_{\max}^{\alpha-1}-L_{\min}^{\alpha-1}} .
    \label{BrmsPrime}
\end{equation} 

    \section{1D diffusion profile}
\label{appendix:diffusion-profile}

In the text, for power estimation in the diffusion regime, we used a 1D profile of the diffusion equation with constant source emission. If the particles' lifetime is given by $T$ and the source is steady, the profile of a particle population is given by
\begin{equation}
    n(s) = Q\int_{t-T}^{t}\dfrac{1}{\sqrt{4\pi D t'}}e^{-x^2/4Dt'} dt' = \int_{0}^{T} ...,
\end{equation}
where the second transformation is done due to temporal invariance of the system.

Evaluating the integral, we obtained the profile~\eqref{eq:1D-diffusion-profile}, which has an additional factor of 2 due to the mirror at $s=0$.

The number of particles between $0 < s < S$ is given by $\int_0^s n(s)\,ds$, the total number of particles is $4QT$. Their fraction $f_e$ is given by
\begin{equation}
    f_e = \rm{erf}(\zeta) + 2\zeta\left[e^{-\zeta}/\sqrt{\pi} - \zeta\rm{erfc}(\zeta) \right],
\end{equation}
where $\zeta = s / \lambda = s / \sqrt{4DT}.$   

    \section{Pitch-angle evolution in non-uniform magnetic fields with losses opposite to momentum}
\label{appendix:guiding-center}

For electrons of $E < 1~\PeV$ in magnetic fields $B\gtrsim 1~\uG$, the cyclotron radius $\rho_c \approx 1~\pc \times E_{15}~(B_{-6}\sin\alpha)^{-1}$ is significantly smaller than the correlation length of magnetic field ($\sim 100~\pc$). This justifies the use of the \emph{guiding center approximation}.

Within this approximation, in the absence of radiative losses, two invariants exist: the exact conservation of energy, $E = \gamma mc^2$, and the adiabatic conservation of the relativistic magnetic moment, $M = \gamma^2 m v_\bot^2 / 2B$~\cite{Northrop1963, Medvedev2026},
\begin{equation}
    d\ln\gamma/dt = 0;\quad d\ln M/dt = 0.
    \label{eq:no-loss-invariants}
\end{equation}

In the ultrarelativistic limit (UrL; $\gamma\gg1$), $d \ln \beta / dt \simeq \gamma^{-2} d\ln\gamma / dt$. Therefore, to leading order, variations of $\beta$ can be neglected and it can be set to unity throughout the equations.

First, this implies conservation of the total momentum, $p^2 = p_\prl^2 + p_\perp^2 \simeq E^2/c^2$ and allows the derivation of the mirror forces acting on the momentum components parallel and perpendicular to the magnetic field, $p_\prl$ and $p_\perp$, respectively:
\begin{equation}
    \dfrac{d p_\prl}{dt}\bigg|_\rm{mag} = - \dfrac{M}{\gamma} \dfrac{dB}{ds};
    \quad
    \dfrac{d p_\perp}{dt}\bigg|_\rm{mag} = \dfrac{Mmv_\prl}{p_\perp}\dfrac{dB}{ds},
    \label{eq:magnetic-forces}
\end{equation}
where $ds = v_\prl dt$ is the differential distance along the magnetic field line.

Second, in the UrL, reaction forces from both synchrotron radiation and inverse Compton scattering are opposite to momentum $\v{F}_\rad = -F_\rad~\v{p}/p$~\cite{Petrosian1985, Singal2016}. In this case, both invariants~\eqref{eq:no-loss-invariants} are broken~\cite{Medvedev2026}. The radiative force components are
\begin{equation}
    \dfrac{d p_\prl}{dt}\bigg|_\rad = -F_\rad\dfrac{p_\prl}{p};
    \quad
    \dfrac{d p_\perp}{dt}\bigg|_\rad = -F_\rad\dfrac{p_\perp}{p}.
    \label{eq:radiative-forces}
\end{equation}

In the UrL, $F_\rad = P_\rad / c$. Following~\cite{Medvedev2026}, we combine the magnetic forces~\eqref{eq:magnetic-forces} with the radiative forces~\eqref{eq:radiative-forces}, to obtain the evolution equations for the particle energy and the relativistic magnetic moment:
\begin{equation}
    \dfrac{d\ln\gamma}{dt} = -\dfrac{P_\rad}{\gamma mc^2};
    \quad
    \dfrac{d\ln M}{dt} = -2\dfrac{P_\rad}{\gamma mc^2}.
\end{equation}

These relations imply the adiabatic conservation of the \emph{classical} magnetic moment, $mv_\perp^2/2B$.  Therefore, in the UrL, the evolution of the particle energy decouples from the pitch-angle evolution driven by the magnetic-field geometry.

Denoting $\mu = \cos\alpha$, we write down the equations describing particle's motion along a magnetic field line
\begin{equation}
    \dfrac{ds}{dt} = \mu c;
    \quad
    \dfrac{d\mu}{dt} = -\mathcal{I}_0\dfrac{dB}{ds};
    \quad
    \dfrac{dE}{dt} = -P_\rad,
    \label{eq:final-evolution-app}
\end{equation}
with $\mathcal{I}_0 = (1-\mu_0^2)/2B_0 = (1-\mu^2)/2B = const$, where $\alpha_0$ and $B_0$ are pitch-angle and magnetic field at the source, respectively.
    \section{Simplified electron transport}
\label{appendix:simplified_transport}

We describe the steady-state electron distribution along the magnetic field line with $n(s,\mu,E) = dN/ds\,d\mu\,dE$. The source of electrons $q$ is point-like and located at $s=0$, hence $q(s,\mu,E) = Q(\mu,E)\delta(s)$.

If we neglect pitch-angle diffusion, the electron transport equation with $\v{u} = (s,\mu,E)$ reads $\p_{\v{u}}(n\dot{\v{u}}) = q(\v{u})$.

It can be solved by method of characteristics with characteristic equations given by~\eqref{eq:final-evolution-app}. By denoting characteristic's parameter $\tau$ and implying zero initial conditions $n(\tau=0) = 0$, we obtain
\begin{equation}
    n(\tau) = \intop_{0}^{\tau} d\tau'~q(\tau') \exp\left\{ -\int_{\tau'}^{\tau} d\tilde\tau~ f(\tilde\tau) \right\},
    \label{eq:general-solution}
\end{equation}
where $f(\tau) = \p_s\dot{s}(\tau) + \p_\mu\dot{\mu}(\tau) + \p_E\dot{E}(\tau) = \p_E\dot{E}(\tau)$.

As the source is point-like, \eqref{eq:general-solution} can be simplified:
\begin{equation}
    n(\tau) = \dfrac{Q(\mu_0, E_0)}{|\mu_0 c|}
    \exp\left\{ -\int_{0}^{\tau} d\tilde\tau~ \p_E\dot{E}(\tilde{\tau}) \right\},
    \label{eq:point-source-simplification}
\end{equation}
where $s(\tau=0) = 0$, $\mu_0 = \mu(0)$,  $E_0 = E(0)$.

We are solving an inverse problem: given $(s, \mu_\rm{obs}, E)$, we want to determine possible initial values $(\mu_0, E_0)$ of the particles getting out from the source.

As seen from~\eqref{eq:general-solution}, pitch-angle evolution is not monotonous, but is energy-independent. Energy evolution is monotonous. Hence, to solve the inverse problem we first define possible injection angle $\mu_0(s) = \pm\sqrt{1 - 2I_0B(s)}$, which could lead to the observed angle $\mu_\rm{obs}$ at position $s$ along the magnetic field line, and secondly calculate backward energy evolution along the selected trajectory.

\begin{figure}[t!]
    \centering
    \includegraphics[width=.8\linewidth]{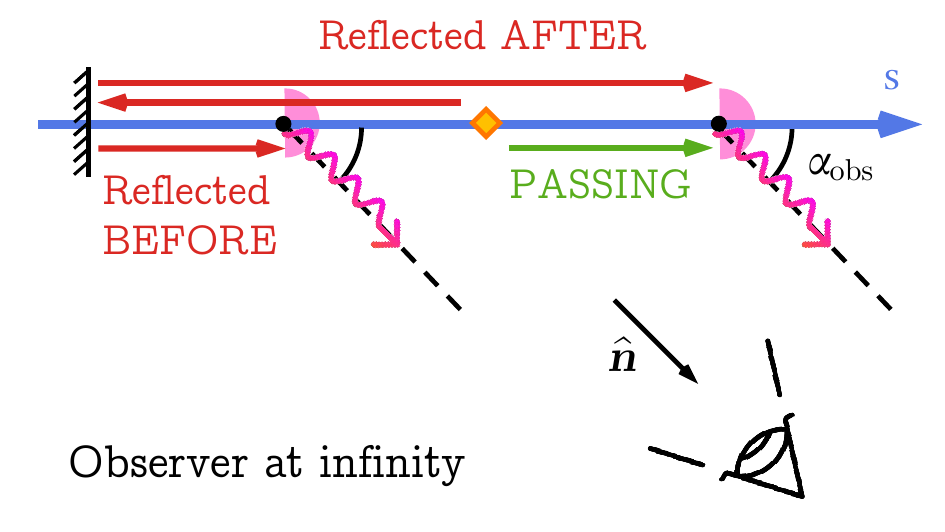}
    \vspace{-3ex}
    \caption{Illustration of particle classification. The source is shown as an orange diamond in the center of a blue magnetic field line. Observer position at infinity is given by vector $\hat{\v{n}}$.}
    \label{fig:passing-reflected}
\end{figure}

In the absence of pitch-angle diffusion, we can classify particles into three groups, dependent on the observed position ($s$, $\mu_\rm{obs}$) and emission direction $\mu_0$. We illustrate this classification in~\autoref{fig:passing-reflected}.

\begin{enumerate}
    \item \emph{passing} particles propagate directly from the source to the observation point, conserving $\mathrm{sign}(\mu)$ along the trajectory. Contribution of these particles to the final emitting electron density is given by 
    \begin{equation}
        n_\rm{P}(s) = \dfrac{Q(\mu_0, E_{0,\rm{P}})}{|\mu_0 c|}
        \exp\left\{ -\int_{0}^{s} \dfrac{d\tilde{s}}{|\mu c|}~ \p_E\dot{E} \right\}.
        \label{eq:point-source-passing}
    \end{equation}

    \item \emph{reflected} particles reach the observation point after changing their direction once at a mirror point. Their contribution to the final electron density depends on the mirror position $s_m$, and is defined with a similar equation to~\eqref{eq:point-source-passing}, where $\int_{0}^{s}$ is substituted by $\int_{0}^{s_m} + \int_{s_m}^{s}$. To account for both \emph{passing} and \emph{reflected}, we introduce notation $\int_{0\to s}$, which accounts for particle trajectories coming from $0$ to $s$.
    
    Based on the relative positions of the mirror point and the observation point with respect to the source, we divide \emph{reflected} particles into two subclasses:
    
    \begin{enumerate}[label=\textit{\roman*}), leftmargin=1.0em, nosep]
        \item \emph{reflected before}: the observation point and the mirror point are on the same side of the source.

        \item \emph{reflected after}: the observation point and the mirror point are on opposite sides of the source.
    \end{enumerate}

    \item \emph{trapped} particles are confined between two mirror points. In UrL, these particles remain trapped indefinitely. We neglect the contribution of these particles, due to cooling. The direct and first reflected passage are described by \emph{passing} and \emph{reflected} particles, respectively.
\end{enumerate}

Each point on the MF line can host particles from different populations. For example, in~\autoref{fig:passing-reflected}, at the right point, electron density consists of both \emph{passing} and \emph{reflected} particles. In our calculation we neglect the contribution of the \emph{trapped} particles as we consider cooling effects to dominate.

    \section{Effective electron cooling}
\label{appendix:effective-cooling}

High-energy electrons in the Galactic Halo are loosing energy via to two dominant mechanisms: by synchrotron radiation in the GMF and by inverse Compton (IC) on CMB photons. Loss rate is given by
\begin{equation}
    -\dot{E} = P_\rm{IC} + A_\rm{syn} E^2 B^2\sin^2\alpha,
    \label{eq:standard_cooling_equation}
\end{equation}
where $A_\rm{syn} = \dfrac{2e^4c}{3(m_ec^2)^4} \simeq 0.12\dm{PeV^{-1}\,kyr^{-1}\,{\uG}^{-2}}$ is the synchrotron cooling constant~\cite{BlumenthalGould1970}.

In the ballistic scenario, we are interested in energy evolution along the electron trajectory, hence we focus on computation of
\begin{equation}
    -\dfrac{dE}{ds} = \dfrac{1}{\mu c}P_\rm{IC} + A_\rm{syn}E^2\dfrac{B^2(1-\mu^2)}{\mu c}
    \label{eq:exact_cooling_equation}
\end{equation}

To accelerate the computations, we define effective cooling time of the particle $t_\rm{c}$ and effective magnetic field $B_\rm{ef}$:
\begin{equation}
    t_\rm{ef} = \intop_\rm{0\to s}\dfrac{d\tilde{s}}{\mu c};
    \quad
    B_\rm{ef}^2 = \intop_\rm{0\to s}\dfrac{d\tilde{s}}{\mu c}B^2(1-\mu^2).
\end{equation}

These definitions allow for simplified treatment of particle cooling, as if it experiences IC and synchrotron losses in constant magnetic field $B_\rm{ef}$ for time $t_\rm{ef}$. Initial energies and ``evolution factors'' (in braces in eq.~\eqref{eq:point-source-simplification}) are precomputed from numerical solutions of~\eqref{eq:standard_cooling_equation} for various values of $B_\rm{c} = B\sin\alpha$ from 0 to $10.0~\uG$. Above $10~\uG$, we neglect IC losses and utilize an analytical solution.

We compared our simplified method a numerical solution of the continuous cooling equation~\eqref{eq:exact_cooling_equation}. The difference in initial energy and the evolution factor are less than~$1\%$ in all energy range of interest in moderate MF and less than $10\%$ above $100~\TeV$ in strong total MF.

\end{document}